\documentclass[aps,prc,reprint,groupedaddress]{revtex4-1}
\usepackage{graphicx} 
\usepackage{epstopdf}
\usepackage{dcolumn} 
\usepackage{bm}
\usepackage{hyperref}
\usepackage{longtable}

\makeatletter

\newcommand{\Rmnum}[1]{\expandafter\@slowromancap\romannumeral #1@}
\makeatother

\begin{document}

\title{Production of proton-rich nuclei around Z=84-90 in fusion-evaporation reactions}

\author{Peng-Hui Chen$^{1,2}$}
\author{Zhao-Qing Feng$^{1}$} \email{Corresponding author:  fengzhq@impcas.ac.cn}
\author{Fei Niu$^{1,3}$}
\author{Ya-Fei Guo$^{1,2}$}
\author{Hong-Fei Zhang$^{2}$}
\author{Jun-Qing Li$^{1}$}
\author{Gen-Ming Jin$^{1}$}

\affiliation{$^{1}$ Institute of Modern Physics, Chinese Academy of Sciences, Lanzhou 730000, People's Republic of China   \\
$^{2}$ School of Nuclear Science and Technology, Lanzhou University, Lanzhou 730000, People's Republic of China
    \\
$^{3}$ Institute of Particle and Nuclear Physics, Henan Normal University, Xinxiang 453007, People's Republic of China
}

\date{\today}
\begin{abstract}
Within the framework of the dinuclear system model, production cross sections of proton-rich nuclei with charged numbers of Z=84-90 are investigated systematically. Possible combinations with the $^{28}$Si, $^{32}$S,  $^{40}$Ar bombarding the target nuclides $^{165}$Ho, $^{169}$Tm, $^{170-174}$Yb, $^{175,176}$Lu, $^{174,176-180}$Hf and $^{181}$Ta are analyzed thoroughly. The optimal excitation energies and evaporation channels are proposed to produce the proton-rich nuclei. The systems are feasible to be constructed in experiments. It is found that the neutron shell closure of N=126 is of importance during the evaporation of neutrons. The experimental excitation functions in the $^{40}$Ar induced reactions can be nicely reproduced. The charged particle evaporation is comparable with neutrons in cooling the excited proton-rich nuclei, in particular for the channels with $\alpha$ and proton evaporation. The production cross section increases with the mass asymmetry of colliding systems because of the decrease of the inner fusion barrier. The channels with pure neutron evaporation depend on the isotopic targets. But it is different for the channels with charged particles and more sensitive to the odd-even effect.

\begin{description}
\item[PACS number(s)]
25.70.Jj, 24.10.-i, 25.60.Pj, 24.60.-k
\end{description}
\end{abstract}

\maketitle

\section{Introduction}

The synthesis of superheavy nuclei (Z$\geq$106) is a very important subject in nuclear physics, motivated by reaching the island of stability predicted theoretically and by exploring the shell evolution and structure properties. It has obtained much progress in experiments with the fusion-evaporation reaction mechanism, i.e., the hot fusion reactions with the $^{48}$Ca bombarding the actinide nuclides and the cold fusion reactions based on the targets of $^{208}$Pb or $^{209}$Bi \cite{Ho00,Og15,Du10,Kh14,Mo07,Gan12}. The heavy proton-rich nuclei (Z$>$83) undergo the $\alpha$ decay and can be easily identified in experiments. The structure properties of the proton-rich nuclei (PRN) associated with the fission barrier, density profiles of neutrons and protons, proton-drip line, level spectra etc would be helpful to extend the superheavy region. A number of models have been developed for understanding the formation mechanism of superheavy nuclei and heavy PRN in the fusion-evaporation reactions \cite{Fe06,Fe09,Fe10,Ba15,Wn11,Wg12,Ad04,Ad14,Ho15,Li12,Lu11,Ro13,Si12,So07}. The formation of superheavy nuclei in the massive fusion reactions is hindered due to the quasifission process. The investigation of the formation mechanism of PRN around Z = 84-90 provides the possibility for exploring the fusion hindrance and can be extrapolated to the superheavy region, which may be useful for synthesizing the new isotopes of heaviest nuclei in the asymmetry actinide-based fusion-evaporation reactions with the emission of charge particles \cite{He16}. Proton-rich nuclei produced in hot fusion reactions around Z = 84-90 are close to the shell closure of N = 126. The survivability of heavy fissile nuclei strongly depends on the fission barrier of a compound nucleus. The barrier is related to the angular momentum and excitation energy \cite{Ar86,An09}. On the other hand, the proton-rich nuclei around Z=84-90 are useful in understanding the decay modes of excited heavy nuclei, which would be helpful for accurately estimating the survival probability of superheavy nuclei in the fusion-evaporation reactions.

In this work, we will investigate the reaction mechanism for producing the proton-rich isotopes of Po, At, Rn, Fr, Ra, Ac, Th in the fusion-evaporation reactions within the dinuclear system (DNS) model \cite{Fe09}. The evaporation residue excitation functions in the $xn$, $xn1p$ and $xn1p1\alpha$ channels with the $x$ being the neutron numbers are analyzed systematically. The article is organized as follows. In section \Rmnum{2} we give a brief description of the DNS model. Calculated results and discussion are presented in section \Rmnum{3}. Summary is concluded in section \Rmnum{4}.

\section{Model description}

The dinuclear system (DNS) is a molecular configuration of two touching nuclei which keep their own individuality.
Such a system has an evolution along two main degrees of freedom: (i) the relative motion of the nuclei in the
interaction potential to form the DNS and the decay of the DNS (qusai-fission process) along the elongation degree of freedom (internuclear motion); (ii) the transfer of nucleons in the mass asymmetry coordinate
$\eta = (A_1 - A_2)/(A_1 + A_2)$ between two nuclei, which is a diffusion process of the excited systems leading to the compound nucleus formation \cite{Fe09}. Off-diagonal diffusion in the surface ($A_1, R$) is not considered since we
assume the DNS is formed at the minimum position of the interaction potential of two colliding nuclei. In this concept, the evaporation residue cross section is expressed as a sum over partial waves with angular momentum
$J$ at the center-of-mass energy $E_{c.m.}$,
\begin{eqnarray}
\sigma^s_{ER}(E_{c.m.}) =&&\frac{\pi\hbar^2}{2\mu E_{c.m.}} \sum \limits ^{J_{max}}_{J=0} (2J+1)T(E_{c.m.},J) \nonumber  \\  && \times P_{CN}(E_{c.m.},J) W^s_{sur}(E_{c.m.},J).
\end{eqnarray}
Here, $T(E_{c.m.})$ is the penetration probability of the two colliding nuclei overcoming the Coulomb barrier to form the DNS, which is calculated using the empirical coupled channel model  \cite{Fe06}. The $P_{CN}$ is the probability that the heavy system evolves from a touching configuration into the formation of compound nucleus in competition with quasi-fission and fission of the heavy fragments  \cite{Fe09}. The last term is the survival probability of the formed compound nucleus, which can be estimated with the statistical evaporation model by considering the competition between neutron evaporation ,$\gamma$-emitting, charged-particle evaporation and fission. we take the maximal angular momentum as $J_{max}$ = 30-50 $\hbar$ since the fission barrier of the heavy nucleus disappears at high spin \cite{Sh10}.

\subsection{Capture cross section and penetration probability}

The capture cross section is given as
\begin{eqnarray}
\sigma_{cap}(E_{c.m.}) = \frac{\pi\hbar^2}{2\mu E_{c.m.}} \sum \limits ^{J_{max}}_{J=0} (2J+1)T(E_{c.m.,J})
\end{eqnarray}
where the penetration probability $T(E_{c.m.},J)$ is evaluated by the Hill-Wheeler formular \cite{Hi53} and the barrier distribution approach.
\begin{eqnarray}
&& T(E_{c.m.},J) = \int f(B) \nonumber \\ &&
\frac{1}{1 + exp(- \frac{2\pi}{\hbar\omega(J)} [ E_{c.m.} - B - \frac{\hbar^2J(J+1)}{2\mu R^2_B(J)}] )} dB
\end{eqnarray}
Here $\hbar\omega(J)$ is the width of the parabolic barrier at the position $R_B(J)$, and the barrier distribution function is assumed to an asymmetric Gaussian form as \cite{Fe06,Za01}
\begin{eqnarray}\label{bdis}
 f(B) = \left\{\begin{array} {rl} \frac{1}{N}exp[-(\frac{B-B_m}{\Delta_1})] & B < B_m \\ \\
 \frac{1}{N}exp[-(\frac{B-B_m}{\Delta_2})] & B > B_m \end{array} \right.
\end{eqnarray}
Here $\Delta_2 = (B_0 -B_s)/2$ ,$\Delta_1=\Delta_2 -2$ MeV, $B_m=(B_0+B_s)/2$£¬$B_0$ and $B_s$ are the height of the barrier at waist-to-waist orientation and the dynamical barrier at the minimal value with varying the quadrupole deformation, respectively. The normalization constant is satisfied to the relation $\int f(B) d B = 1$. The nucleus-nucleus interaction potential is given by
\begin{eqnarray}
V(R;\beta_1,\beta_2,\theta_1,\theta_2) = V_C(R;\beta_1,\beta_2,\theta_1,\theta_2) +     \nonumber \\  V_N(R;\beta_1,\beta_2, \theta_1,\theta_2) + \frac{1}{2}C_1(\beta_1-\beta^0_1)^2 + \frac{1}{2}C_2(\beta_2-\beta^0_2)^2     \nonumber \\
\end{eqnarray}
Here the numbers 1 and 2 denote the projectile and the target, respectively. The $\beta_{1,2}$ are the parameters of the dynamic quadrupole deformation, and $\beta^0_{1,2}$ are the parameters of static deformation. To reduce the number of variables we assume that the deformation energy of two nuclei are proportional to their mass \cite{Za01}, i.e., $C_1\beta^2_1 / C_2\beta^2_2 = A_1/A_2$, and we may use only one deformation parameter $\beta = \beta_1 + \beta_2$ and the $C_i$ (i=1,2) are the stiffness parameters, which were calculated within the liquid drop model \cite{My66}
\begin{eqnarray}
C_i = (\lambda - 1) ((\lambda - 1)R^2_i\sigma - \frac{3}{2\pi}\frac{Z^2e^2}{R_i(2\lambda + 1)})
\end{eqnarray}
Where $R_i$ is the radius of the nucleus. Here, we only take into account the quadrupole deformation ($\lambda = 2$). $\sigma$ is the coefficient of surface tension which satisfies $4\pi R^2_i \sigma = a_s A^{2/3}_i$, and $a_s$ = 18.32 MeV is the surface energy.

The nuclear potential is calculated by the double-folding method based on the Skyrme interaction force without considering the momentum and the spin dependence as \cite{Ad96,Fe09}
\begin{eqnarray}\label{dfp}
&& V_N = C_0 \left\{ \frac{F_{in} - F_{ex}}{\rho_0} \left[ \int \rho_1^2(\textbf{r}) \rho_2(\textbf{r}-\textbf{R}) dr \right. \right. \nonumber \\ && \left. \left. + \int \rho_1(\textbf{r}) \rho_2^2(\textbf{r}-\textbf{R}) d\textbf{r} \right] + F_{ex}\int \rho_1(\textbf{r}) \rho_2(\textbf{r}-\textbf{R}) d\textbf{r} \right\} \nonumber \\
\end{eqnarray}
with
\begin{eqnarray}
F_{in,ex} = f_{in,ex} + f'_{in,ex}\frac{N_1 - Z_1}{A_1}\frac{N_2 - Z_2}{A_2}
\end{eqnarray}
which is dependent on the nuclear densities and on the orientations of deformed nuclei. The parameters $C_0$ = 300 MeV fm$^3$, $f_{in}$ = 0.09, $f_{ex}$ = -2.59, $f_{in}'$ = 0.42, $f_{ex}'$ = 0.54, and $\rho_0$ = 0.16 fm$^3$ are used in the calculation. The Woods-Saxon density distributions are expressed for two nuclei as
\begin{eqnarray}
\rho_1(\textbf{r}) = \frac{\rho_0}{1 + exp[(\textbf{r} - \Re_1(\theta_1))/a_1]}
\end{eqnarray}
and
\begin{eqnarray}
\rho_2(\textbf{r} - \textbf{R}) = \frac{\rho_0}{1 + exp[(|\textbf{r} - \textbf{R}| - \Re_2(\theta_2))/a_2]}
\end{eqnarray}
Here $\Re_i$ ($\theta_i$) (i = 1, 2) are the surface radii of the nuclei with $\Re_i$ ($\theta_i$) = $R_i[1 + \beta_iY_{20}(\theta_i)]$, and the spheroidal radii are $R_i$ . The parameters $a_i$ represent the surface diffusion coefficients, which are taken as 0.55 fm in the calculation.In the actual calculation, the distance \textbf{R} between the centers of the two fragments is chosen to be the value that gives the minimum of the interaction potential, in which the DNS is considered to be formed.

The Coulomb potential is obtained by Wong's formula \cite{Wo73}, which agrees well with the double-folding procedure.
\begin{eqnarray}\label{cp}
V_C(r,\beta_1,\beta_1,\theta_1,\theta_2) = \frac{Z_1Z_2e^2}{r} + (\frac{9}{20\pi})^{1/2}(\frac{Z_1Z_2e^2}{r^3}) \sum \limits _{i=1}^2R_i^2 \beta_i \nonumber \\ P_2(\cos\theta_i) + (\frac{3}{7\pi})(\frac{Z_1Z_2e^2}{r^3})\sum \limits _{i=1}^2R_i ^2[\beta_i P_2(\cos\theta_i)]^2  \nonumber \\
\end{eqnarray}
Where $\theta_i$, $\beta_i$, $R_i$, and $P_2(\cos\theta_i)$ are the angle between the symmetry axis of the deformed projectile or target and the collision axis, quadrupole deformation parameter, the radius of projectile or target, the Legendre polynomial, respectively.

\subsection{Fusion probability}

In order to describe the fusion dynamics as a diffusion process in mass asymmetry, the analytical solution of the Fokker-Planck equation and the numerical solution of the master equations have been used, which were also used to treat deep inelastic heavy-ion collisions. Here, the fusion probability is obtained by solving a set of master equations numerically in the potential energy surface of the DNS. The time evolution of the distribution function $P(A_1, E_1, t)$ for fragment 1 with mass number $A_1$ and excitation energy $E_1$ is described by the following master equations \cite{Fe09,Fe10},
\begin{eqnarray}
&&\frac{d P(Z_1,N_1,E_1,t)}{d t} = \nonumber \\ && \sum \limits_{Z'_1}W_{Z_1,N_1;Z_1,N'_1}(t) [d_{Z_1,N_1}P(Z'_1,N_1,E'_1,t) \nonumber \\ && - d_{Z'_1,N_1}P(Z_1,N_1,E_1,t)] + \nonumber \\ &&
 \sum \limits_{N'_1}W_{Z_1,N_1;Z_1,N'_1}(t)[d_{Z_1,N_1}P(Z_1,N'_1,E'_1,t) \nonumber \\ && - d_{Z_1,N'_1}P(Z_1,N_1,E_1,t)] - \nonumber \\
 &&[\Lambda ^{qf}_{A_1,E_1,t}(\Theta) + \Lambda^{fis}_{A_1,E_1,t}(\Theta)]P(A_1,E_1,t).
\end{eqnarray}
Here the $W_{Z_1,N_1,Z'_1,N_1}$ ($W_{Z_1,N_1,Z'_1,N_1}$) is the mean transition probability from the channel ($Z_1,N_1,E_1$) to ($Z'_1,N_1,E'_1$) [or ($Z_1,N_1,E_1$) to ($Z_1,N'_1,E'_1$)], and $d_{Z_1,N_1}$ denotes the microscopic dimension corresponding to the macroscopic state ($Z_1,N_1,E_1$). The sum is taking all possible proton and neutron numbers that fragment $Z'_1$, $N'_1$ may take, but only one nucleon transfer is considered in the model with the relation $Z'_1$ = $Z_1$ $\pm$ 1, and $N'_1$ = $N_1$ $\pm$ 1. The excitation energy $E_1$ is the local excitation energy $\varepsilon^*_1$ with respect to fragment $A_1$, which is determined by the dissipation energy from the relative motion and potential energy surface of DNS \cite{Li83,Wo78}. The dissipation energy is described by the parametrization method of classical deflection function \cite{Li81}. The motion of nucleons in the interacting potential is governed by the single-particle Hamiltonian \cite{Wn75,Fe87} as
\begin{eqnarray}
H(t) = H_0(t) + V(t)
\end{eqnarray}
with
\begin{eqnarray}
H_0(t) && = \sum _K\sum_{\nu_K} \varepsilon_{\nu_K}(t)\alpha^+_{\nu_K}(t)\alpha_{\nu_K}(t) \nonumber \\
V(t) && = \sum_{K,K^{'}} \sum_{\alpha_K,\beta_{K'}} u_{\alpha_K,\beta_{K'}}\alpha^+_{\alpha_K}(t)\alpha_{\beta_K}(t)  \\ && = \sum_{K,K'}V_{K,K'}(t) \nonumber
\end{eqnarray}
Here the indices $K$, $K'$ ($K,K'$ = 1, 2) denote the fragment 1 and 2. The quantities $\varepsilon_{\nu K}$ and $u_{\alpha_K,\beta_{K'}}$ represent the single particle energies and the interaction matrix elements, respectively. The single particle states are defined with respect to the centers of the interacting nuclei and are assumed to be orthogonalized in the overlap region. So the annihilation and creation operators are dependent on time. The single particle matrix elements are parameterized by
\begin{eqnarray}
u_{\alpha_K,\beta_K'} = && U_{K,K'}(t)  \\ && \left\{ exp \left[- \frac{1}{2}( \frac{\varepsilon_{\alpha_K}(t) - \varepsilon_{\beta_K}(t)}{\Delta_{K,K'}(t)})^2 \right] - \delta_{\alpha_K,\beta_{K'}} \right\}  \nonumber
\end{eqnarray}
which contain some parameters $U_{K,K'}(t)$ and $\delta_{\alpha_K,\beta_{K'}}(t)$. The detailed calculation of these parameters and the mean transition probabilities were described in Ref. \cite{Wn75}.

The evolution of the DNS along the variable $R$ leads to the quasi-fission of the DNS. The quasi-fission rate $\Lambda^{qf}$ can be estimated with the one-dimensional Kramers formula \cite{Ad03,Pg83}:
\begin{eqnarray}\label{qf}
\Lambda^{qf}_{A_1,E_1,t}(\Theta) = && \frac{\omega}{2\pi\omega^{B_{qf}}}\left[\sqrt{(\frac{\Gamma}{2\hbar})^2 + (\omega^{B_{qf}})^2} - \frac{\Gamma}{2\hbar}\right]            \nonumber      \\
&&  \exp\left(- \frac{B_{qf}(A_1)}{\Theta(A_1,E_1,t)}\right)
\end{eqnarray}
Here the quasi-fission barrier is counted from the depth of the pocket of the interaction potential. The local temperature is given by the Fermi-gas expression $\Theta = \sqrt{\varepsilon^*/a}$ corresponding to the local excitation energy $\varepsilon^*$ and level density parameter $a$ in Eq. (\ref{pld}). In Eq. (\ref{qf}) the frequency $\omega^{B_{qf}}$ is the frequency of the inverted harmonic oscillator approximating the interaction potential of two nuclei in $R$ around the top of the quasi-fission barrier, and $\omega$ is the frequency of the harmonic oscillator approximating the potential in $R$ around the bottom of the pocket. The quantity $\Gamma$, which denotes the double average width of the contributing single-particle states, determines the friction coefficients: $\gamma_{ii'} = \frac{\Gamma}{\hbar}\mu_{ii'}$, with $\mu_{ii'}$ being the inertia tensor. Here we use constant values $\Gamma$ = 2.8 MeV, $\hbar \omega^{B_{qf}}$ = 2.0 MeV and $\hbar \omega$ = 3.0 MeV for the following reactions. The Kramers formula is derived with the quasi-stationary condition of the temperature $\Theta(t)$ $\leq$ $B_{qf}(A1,A2)$. However, the numerical calculation in Ref. \cite{Pg83} indicated that Eq. (\ref{qf}) is also useful for the condition of $\Theta(t)$ $>$ $B_{qf}(A1,A2)$. In the reactions of synthesizing superheavy nuclei or PRN, there is the possibility of the fission of the heavy fragment in the DNS. Because the fissility increases with the charge number of the nucleus, the fission of the heavy fragment can affect the quasi-fission and fusion when the DNS evolves towards larger mass asymmetry. The fission rate $\Lambda^{fis}$ can also be treated with the one-dimensional Kramers formula \cite{Ad03}
\begin{eqnarray}
\Lambda^{fis}_{A_1,E_1,t}(\Theta) = && \frac{\omega_{g.s}}{2\pi\omega_f}\left[\sqrt{(\frac{\Gamma_0}{2\hbar})^2 + (\omega_f)^2} - \frac{\Gamma_0}{2\hbar}\right]   \nonumber       \\
 &&            \exp\left(- \frac{B_f(A_1)}{\Theta(A_1,E_1,t)}\right)
\end{eqnarray}
where the $\omega_{g.s.}$ and $\omega_f$ are the frequencies of the oscillators approximating the fission-path potential at the ground state and on the top of the fission barrier for nucleus $A_1$ or $A_2$ (larger fragment), respectively. Here, we take $\hbar\omega_{g.s.}$ = $\hbar\omega_f$ = 1.0 MeV, $\Gamma_0$ = 2 MeV. The fission barrier is calculated as the sum of a macroscopic part and the shell correction energy (see Eq. (\ref{fb})). The fission of the heavy fragment does not favor the diffusion of the system to a light fragment distribution. Therefore, it leads to a slight decrease of the fusion probability (see\ Eq. (\ref{pcn})).

In the relaxation process of the relative motion, the DNS will be excited by the dissipation of
the relative kinetic energy. The excited system opens a valence space $\Delta \varepsilon_K$ in fragment $K$ ($K$ = 1, 2), which has a symmetrical distribution around the Fermi surface. Only the particles in the
states within this valence space are actively involved in excitation and transfer. The averages on
these quantities are performed in the valence space:
\begin{eqnarray}
\Delta \varepsilon_K = \sqrt{\frac{4\varepsilon^*_K}{g_K}},\quad
\varepsilon^*_K =\varepsilon^*\frac{A_K}{A}, \quad
g_K = A_K /12,
\end{eqnarray}
where the $\varepsilon^*$ is the local excitation energy of the DNS, which provides the excitation energy
for the mean transition probability. There are $N_K$ = $g_K\Delta\varepsilon_K$ valence states and $m_K$ = $N_K/2$ valence nucleons in the valence space $\Delta\varepsilon_K$, which gives the dimension
\begin{eqnarray}
 d(m_1, m_2) = {N_1 \choose m_1} {N_2 \choose m_2}.
\end{eqnarray}
The local excitation energy is defined as
\begin{eqnarray}
\varepsilon^* = E_x - (U(A_1)-U(A_P))
\end{eqnarray}
Here the $U(A_1, A_2)$ and $U(A_P, A_T)$ are the driving potentials of fragments $A_1$, $A_2$ and fragments
$A_P$, $A_T$ (at the entrance point of the DNS), respectively. The detailed calculation of the driving potentials is from  Eq. (\ref{dp}). The excitation energy $E_x$ of the composite system is converted from the relative kinetic energy loss, which is related to the Coulomb barrier $B$ \cite{Fe07}.

The potential energy surface (PES; i.e., the driving potential) of the DNS is given by
\begin{eqnarray}\label{dp}
&& U(A_1, A_2, J, \textbf{R}; \beta_1, \beta_2, \theta_1, \theta_2) = B(A_1) + B(A_2) - \nonumber \\ && [ B(A) + V^{CN}_{rot}(J) ] + V(A_1, A_2, J, \textbf{R}; \beta_1, \beta_2, \theta_1, \theta_2)
\end{eqnarray}
with $A_1 + A_2 = A$. Here $B(A_i)(i = 1, 2)$ and $B(A)$ are the negative binding energies of the fragment $A_i$ and the compound nucleus $A$, respectively, in which the shell and the pairing corrections are included reasonably; $V^{CN}_{rot}$ is the rotation energy of the compound nucleus; the $\beta_i$ represent quadrupole deformations of the two fragments; the $\theta_i$ denote the angles between the collision orientations and the symmetry axes of deformed nuclei. The interaction potential between fragment $1(Z_1, A_1)$ and $2(Z_2,A_2)$ includes the nuclear, Coulomb, and centrifugal parts as
\begin{eqnarray}
&& V(A_1, A_2, J, \textbf{R}; \beta_1, \beta_2, \theta_1, \theta_2) = \nonumber \\ && V_N(A_1, A_2, J, \textbf{R}; \beta_1, \beta_2, \theta_1, \theta_2) \nonumber \\ && + V_C(A_1, A_2, J, \textbf{R}; \beta_1, \beta_2, \theta_1, \theta_2) + \frac{J(J+1)\hbar^2}{2\mu \textbf{R}^2}
\end{eqnarray}
where the reduced mass is given by $\mu = m \cdot A_1 A_2 / A$ with the nucleon mass $m$. The nuclear potential and Coulomb potential are taken from Eq. (\ref{dfp}) and Eq. (\ref{cp}), respectively. In the actual calculation, the distance \textbf{R} between the centers of the two fragments is chosen to be the value that gives the minimum of the interaction potential, in which the DNS is considered to be formed. So the PES depends only on the mass asymmetry degree of freedom $\eta$, which gives the driving potential of the DNS.

After reaching the reaction time in the evolution of $P(A_1, E_1, t)$, all those components on the left side of the BG (Businaro-Gallone) point contribute to the formation of the compound nucleus. The hindrance in the diffusion process by nucleon transfer to form the compound nucleus is the inner fusion barrier $B_{fus}$, which is defined as the difference of the driving potential at the BG point and at the entrance position. Nucleon transfers to more symmetric fragments that leads to quasi-fission. The formation probability of the compound nucleus at the Coulomb barrier $B$ and angular momentum $J$ is given by
\begin{eqnarray}
P_{CN}(E_{c.m.} ,J, B) = \sum \limits ^{A_{BG}}_{A_1=1}P(A_1,E_1,\tau_{int}(E_{c.m.},J,B))\nonumber \\
\end{eqnarray}
Here the interaction time $\tau_{int}(E_{c.m.},J,B)$ is obtained using the deflection function method \cite{Li83}, which means the time duration for nucleon transfer from the capture stage to the formation of the complete fusion system with the order of $10^{-20}$ s. We obtain the fusion probability as
\begin{eqnarray}\label{pcn}
P_{CN}(E_{c.m.},J) = \int f(B)P_{CN}(E_{c.m.},J,B)d B
\end{eqnarray}
where the barrier distribution function is taken in asymmetric Gaussian form (see Eq. (\ref{bdis})). So the fusion cross section is written as
\begin{eqnarray}
\sigma_{fus}(E_{c.m.}) = \frac{\pi\hbar^2}{2\mu E_{c.m.}} \sum \limits^{ \infty }_{J=0} (2J+1)T(E_{c.m.,J})P_{CN}(E_{c.m.},J)  \nonumber \\
\end{eqnarray}

\subsection{The survival probability of the excited compound nucleus}

The survival probability is particularly important in evaluation of the cross section, which is usually calculated with the statistical approach. The physical process in understanding the excited nucleus is clear. But the magnitude strongly depends on the ingredients in the statistical model, such as level density, separation energy, shell correction, fission barrier etc. The excited nucleus is cooled by evaporating $\gamma$-rays, light particles (neutrons, protons, $\alpha$ etc) in competition with fission. Similar to neutron evaporation \cite{Fe10}, the probability in the channel of evaporating the $x-$th neutron, the $y-$th proton and the $z-$ alpha is expressed as \cite{Ch16}
\begin{eqnarray}
&&W_{sur}(E^*_{CN},x,y,z,J) =  P(E^*_{CN},x,y,z,J)             \nonumber   \\
&& \times   \prod^x_{i=1} \frac{\Gamma_n(E^*_i,J)}{\Gamma_{tot}(E^*_i,J)}
\prod^y_{j=1} \frac{\Gamma_p(E^*_j,J)}{\Gamma_{tot}(E^*_j,J)}
 \prod^z_{k=1} \frac{\Gamma_{\alpha}(E^*_k,J)}{\Gamma_{tot}(E^*_k,J)}.
\end{eqnarray}
Here the $E^*_{CN}$, $J$ are the excitation energy and the spin of the excited nucleus, respectively. The total width $\Gamma_{tot}$  is the sum of partial widths of particle evaporation, $\gamma$-emission and fission. The excitation energy $E^*_s$ before evaporating the $s$-th particle is evaluated by
\begin{equation}
E^*_{s+1} = E^*_s - B^n_i - B^p_j - B^{\alpha}_k - 2T_s
\end{equation}
with the initial condition $E^*_1 = E^*_{CN}$ and $s=i+j+k$. The $B^n_i$, $B^p_j$, $B^{\alpha}_k$ are the separation energy of the $i$-th neutron, $j$-th proton, $k$-th alpha, respectively. The nuclear temperature $T_i$ is given by $E^*_i = aT_i^2-T_i$ with $a$ being the level density parameter.

Assuming the electric dipole radiation (L=1) dominates $\gamma-$emission, the decay width is calculated by
\begin{equation}
\Gamma_\gamma(E^*_{CN},J) = \frac{3}{\rho(E^*,J)}    \nonumber  \\
\int \limits ^{E^*-\delta -\frac{1}{a}}_{\varepsilon = 0} \rho(E^*-E_{rot}-\varepsilon,J) f_{E_1}(\varepsilon) d \varepsilon,
\end{equation}
and
\begin{equation}
f_{E_1}(\varepsilon) = \frac{4}{3\pi} \frac{1+\kappa}{mc^2} \frac{e^2}{\hbar c} \frac{NZ}{A} \frac{\Gamma_G\varepsilon^4}{(\Gamma_G\varepsilon)^2 + (\Gamma_G^2 - \varepsilon^2)^2}.
\end{equation}
Here, $\kappa = 0.75$, and $\Gamma _G$ and $E_G$ are the width and position of the electric dipole resonance respectively. For a heavy nucleus, $\Gamma _G = 5$ MeV \cite{Sc91},
\begin{equation}
E_G = \frac{167.23}{A^{1/3}\sqrt{1.959 + 14.074A^{-1/3}}}.
\end{equation}

The particle decay widths are evaluated with the Weisskopf evaporation theory as \cite{We37}
\begin{eqnarray}
&& \Gamma_\nu(E^*,J) = (2s_\nu + 1) \frac{m_\nu}{\pi^2 \hbar^2 \rho(E^*,J)} \int \limits ^{E^*-B_\nu-\delta-\delta_n -\frac{1}{a}}_0 \nonumber \\ && \varepsilon \rho(E^*-B_\nu - \delta_n-E_{rot}-\varepsilon,J)\sigma_{inv}(\varepsilon) d \varepsilon .
\end{eqnarray}
Here, $s_\nu$, $m_\nu$ and $B_\nu$  are the spin, mass and binding energy of the evaporating particle, respectively. The pairing correction energy $\delta$ is set to be $12/\sqrt{A}, 0, -12/\sqrt{A}$  for even-even, even-odd and odd-odd nuclei, respectively. The inverse cross section is given by $\sigma_{inv}=\pi R_\nu^{2}T(\nu) $. The penetration probability is set to be unity for neutrons and $T(\nu) =(1 + \exp(\pi(V_C(\nu)-\varepsilon)/\hbar\omega))^{-1}$ for charged particles with $\hbar \omega= 5 $ and 8 MeV for proton and alpha, respectively.

The fission width is calculated with a similar method, which is given by Bohr-Wheeler formula as in Ref. \cite{Fe09,Fe10}.
\begin{eqnarray}
&& \Gamma_f(E^*,J) = \frac{1}{2\pi \rho_f(E^*,J)} \int \limits ^{E^*-B_f-E_{rot}-\delta-\delta_f -\frac{1}{a_f}}_0  \nonumber \\ &&
 \frac{ \rho_f(E^*-B_f - \delta_f-E_{rot}-\varepsilon,J) d \varepsilon}{1+exp[-2\pi(E^*-B_f - \delta_f-E_{rot}-\varepsilon)/\hbar\omega]},
\end{eqnarray}
where $\omega\hbar$ = 2.2 MeV is the width of distribution of fission \cite{As05}, and $\delta_f$ is a correction for fission barrier ( $\delta_f$ = $\delta$ for even-even or odd-odd nucleus, else $\delta$ = 0 ). $B_f$ is the fission barrier , which is mainly determined by the macroscopic part (calculated by liquid drop model) and microscopic shell correction energy and given as
\begin{eqnarray}\label{fb}
B_f(E^*, J) = && B^{LD}_f + B^{M}_{f}(E^* = 0,J)exp(-E^*/E_D) \nonumber \\ && - (\hbar^2/2\zeta_{g.s.}-\hbar^2/2\zeta_{sd})J(J+1)
\end{eqnarray}
where the macroscopic part is calculated by liquid drop model,
\begin{eqnarray}
B^{LD}_f = \left \{\begin{array} {rl} 0.38(0.75 - x )E_{s0} & , (1/3 < x < 2/3) \\ \\
 0.83(1-x)^3 E_{s0} & ,(2/3 < x < 1) \end{array} \right.
\end{eqnarray}
Here fissility parameter $x$ can be given as
\begin{eqnarray}
x = \frac{E_{c0}}{2E_{s0}}.
\end{eqnarray}
Where $E_{s0}$ and $E_{c0}$ are the surface energy of spherical nuclei and coulomb energy, respectively, by the Myers-Swiatecki mass formula \cite{My74} which can be given as
\begin{eqnarray}
 && E_{s0} = 17.944\left[1-1.7826(\frac{N-Z}{A})^2 \right]A^{2/3} \ MeV, \\
 && E_{c0} = 0.7053 \frac{Z^2}{A^{1/3}} \ MeV
\end{eqnarray}
Microcosmic shell correction energy is taken from Ref. \cite{Mo95,Mo15}. Shell damping energy is
\begin{eqnarray}
E_D = \frac{5.48A^{1/3}}{1+1.3A^{-1/3}} MeV \ or \ E_D = 0.4A^{4/3}/a.
\end{eqnarray}
Here $a$ is the parameter of level density (see Eq. (\ref{pld}), for fission level density $a_f$ = 1.1$a$. The moments of inertia of fission compound nuclei in its ground state (g.s.) and at the saddle point configuration (sd) are given as
\begin{eqnarray}\label{im}
\zeta_{g.s.,sd} = k\times \frac{2}{5}MR^2(1+\beta^{g.s.,sd}_2/3)
\end{eqnarray}
where $k = 0.4$ ,$\beta_2$ is the parameter of quadrupole deformation in ground state, which are got from Ref.\cite{Mo95}; $\beta^{sd}_2 = \beta^{g.s.}_2 + 0.2$ is the the parameter of quadrupole deformation in saddle point configuration, which is calculated by relativistic mean field theory \cite{We03}.

The level density is calculated from the Fermi-gas model~\cite{Ig79} as,
\begin{eqnarray}
\rho(E^*,J) = K_{coll}\cdot \frac{2J+1}{24\sqrt{2}\sigma^3a^{1/4}(E^*-\delta)^{5/4}}  \nonumber\\[1mm]
exp\left[ 2\sqrt{a(E^*-\delta)} - \frac{(J+1/2)^2}{2\sigma^2}\right],
\end{eqnarray}
with $\sigma^2 = 6\bar{m}^2\sqrt{a(E^*-\delta)}/\pi^2$  and $\bar{m}\approx0.24A^{2/3}$. The $K_{coll}$ is the collective enhancement factor, which includes the rotational and vibrational effects \cite{Fe10,Ju98}. The level density parameter is related to the shell correction energy $E_{sh}(Z,N)$ and the excitation energy $E^{\ast}$ of the nucleus as
\begin{eqnarray}\label{pld}
a(E^{\ast},Z,N)=\tilde{a}(A)[1+E_{sh}(Z,N)f(E^{\ast}-\Delta)/(E^{\ast}-\Delta)]. \nonumber \\
\end{eqnarray}
Here, $\tilde{a}(A)=\alpha A + \beta  A^{2/3}b_{s}$ is the asymptotic Fermi-gas value of the level density parameter at high excitation energy. The shell damping factor is given by
\begin{eqnarray}
f(E^{\ast})=1-\exp(-\gamma E^{\ast})
\end{eqnarray}
with $\gamma=\tilde{a}/(\epsilon  A^{4/3})$. The parameters $\alpha$, $\beta$, $b_{s}$ and $\epsilon$ are taken to be 0.114, 0.098, 1. and 0.4, respectively \cite{Fe10}. The charged particles (p, $\alpha$) have smaller widths for the superheavy nucleus in comparison to  the proton-rich nucleus because of larger separation energies. The fission width increases rapidly in the excitation energy range of 10 - 30 MeV for the superheavy nucleus and the larger width leads to a smaller survival probability, which is because the fission barrier decreases exponentially with increasing excitation energy ~\cite{Fe10}. The collective enhancement factor increases the level density, but reduces the partial widths, in particular for particle evaporation.

For one particle evaporation, the realization probability is given by
\begin{eqnarray}
P(E^*_{CN},J) = \exp\left( -\frac{(E^*_{CN} - B_s - 2T)^2}{2\sigma^2} \right).
\end{eqnarray}
The width $\sigma$ is taken to fit the experimental width of fusion-evaporation excitation functions. The realization probability $P(E^*_{CN},x,y,z,J)$ for evaporating $x$ neutrons, $y$ protons, $z$ alphas at the excitation energy of $E^*_{CN}$ and angular momentum of $J$ is calculated by the Jackson formula~\cite{Ja56} as
\begin{eqnarray}
P(E^*_{CN},s,J) = I(\Delta_s,2s-3) - I(\Delta_{s+1},2s-1),
\end{eqnarray}
where the quantities $I$ and $\Delta$ are given by following:
\begin{eqnarray}
I(z,m) = \frac{1}{m!} \int ^z_0 u^m e^{-u} d u,   \\
\Delta_s = \frac{E^*_{CN} - \sum \limits ^s_{i=1}B^\nu_i}{T_i}.
\end{eqnarray}
The $B^\nu_i$ is the separation energy of evaporating the $i$-th particle and $s(x,y,z)=x+y+z$. The spectrum of the realization probability determines the structure of survival probability in each evaporation channel.

\section{Results and discussion}

\subsection{Comparison with the experimental data}

TABLE \Rmnum{1}. The calculated evaporation residue cross-sections $\sigma^{th}_{ER}$ and the available experimental data $\sigma^{exp}_{ER}$ \cite{Ve84} in the $^{40}$Ar induced reactions on the targets of $^{165}$Ho, $^{169}$Tm, $^{171,174}$Yb, $^{175}$Lu, $^{176-180}$Hf, and $^{181}$Ta.

\begin{longtable}{ccccc}
\hline\hline
    Target           & Channel & $E_{C.N.}(MeV)$  &   $\sigma^{exp}_{ER}$     &  $\sigma^{th}_{ER}$
 \\
 \hline
 \\
  $ ^{165}Ho$  &  3n  &    40.0    &  4.4 mb $\pm 1.7 \% $      &   0.80 mb
  \\
                        &  4n  &    49.0    &  11.3 mb $ \pm 1. \% $     &   4.64 mb
                        \\
                        &  5n  &    57.0    &   9.8 mb $ \pm 1.7 \% $    &   3.88 mb
                        \\
                        &  6n  &    69.0    &   2.0 mb $ \pm 0.6 \% $     &  2.15  mb
                        \\ 
                        \\
  $ ^{169}Tm$  & 2n+\underline{3n} &   42.0    &  376 $\mu b\ \pm 1.8 \% $   &   75.8 $\mu b$
  \\
                        &  4n   &    50.0    &  545 $\mu b\ \pm 1.2 \% $  &   449.4 $\mu b$
                        \\
                        &  5n   &    59.0    &  155 $\mu b\ \pm 11. \% $  &   540 $\mu b$
                        \\
                        &  6n   &    74.0    &  16 $\mu b\ \pm 11.5 \% $  &   123 $\mu b$
                        \\ 
                        \\
  $ ^{171}Yb$  & 3n+\underline{4n} &    50.0    &  34 $\mu b\ \pm 4.5 \% $   &   2.4 $\mu b$
  \\
                        &  5n   &    63.0    &  4.9 $\mu b\ \pm 45. \% $  &   1.36 $\mu b$
                        \\
                        &  6n   &    76.0    &  3.4 $\mu b\ \pm 39. \% $  &   0.9 $\mu b$
                        \\
                        &p1n+\underline{p2n}&    43.0    &  11.8 $\mu b\ \pm 27. \% $ &   0.97 $\mu b$
                        \\
                        &p3n+\underline{p4n}&    60.0    &  55.3 $\mu b\ \pm 15. \% $ &   42.0 $\mu b$
                        \\ 
                        \\
  $ ^{174}Yb$  & 2n+\underline{3n} &    45.0    &  277 $\mu b\ \pm 5.3 \% $  &   1.9 $\mu b$
  \\
                        & 4n+\underline{5n} &    57.0    &  1.4 $mb\ \pm 2.3 \% $     &   0.94 mb
                        \\
                        & \underline{6n}+7n &    69.0    &  86.7 $\mu b\ \pm 8.5 \% $ &   23.0 $\mu b$
                        \\
                        &p2n+\underline{p3n}&    50.0    &  73.8 $\mu b\ \pm 3.2 \% $ &   17.0 $\mu b$
                        \\
                        &\underline{p4n}+p5n&    68.0    &  228 $\mu b\ \pm 6. \% $   &   120.0 $\mu b$
                        \\ 
                        \\
  $ ^{175}Lu$  & 2n+\underline{3n} &    42.0    &  42.1 $\mu b\ \pm 8.7 \% $ &   10.0 $\mu b$
  \\
                        & 4n+\underline{5n} &    61.0    &  68.8 $\mu b\ \pm 3.5 \% $ &   217.3 $\mu b$
                        \\
                        &  6n   &    73.0    &  2. $\mu b\ \pm 18 \% $    &   26.2 $\mu b$
                        \\
                        &p2n+\underline{p3n}&    48.0    &  24.2 $\mu b\ \pm 3.7 \% $ &   5.2 $\mu b$
                        \\
                        &\underline{p4n}+p5n&    69.0    &  67.6 $\mu b\ \pm 8.1 \% $ &   11.1 $\mu b$
                        \\
        & $\alpha 2n+\underline{\alpha 3n}$ &    48.0    &  112 $\mu b\ \pm 46 \% $   &   18.4 $\mu b$
        \\
        & $\underline{\alpha 4n}+\alpha 5n$ &    69.0    &  246 $\mu b\ \pm 13 \% $   &   81.6 $\mu b$
        \\ 
        \\
  $ ^{176}Hf$  & 2n+\underline{3n} &    43.0    &  530 $ nb\ \pm 46 \% $     &   153.9 $ nb$
  \\
                        & 4n+\underline{5n} &    53.0    &  174 $ nb\ \pm 36 \% $     &   178.7 $ nb$
                        \\
                        &p2n+\underline{p3n}&    50.0    &  3.4 $\mu b\ \pm 10 \% $   &   8.0 $\mu b$
                        \\
                        &\underline{p4n}+p5n&    64.0    &  2.1 $\mu b\ \pm 4.5 \% $  &   6.5 $\mu b$
                        \\
        & $\alpha$2n+\underline{$\alpha$3n} &    49.0    &  24 $\mu b\ \pm 12 \% $    &   12.5 $\mu b$
        \\
        & \underline{$\alpha$4n}+$\alpha$5n &    63.0    &  9.4 $\mu b\ \pm 9. \% $   &   13.5 $\mu b$
        \\
        \\
$ ^{177}Hf$ &    3n+\underline{4n} &    48.0         &  633 $ nb\ \pm 13 \% $       &   876 $ nb$
   \\
                &    \underline{5n}+6n      &    62.0         &  191 $ nb\ \pm 41 \% $       &   275 $ nb$
                \\
                &     p2n       &    41.0         &  899 $ nb\ \pm 17 \% $       &   107 $ nb$
                \\
                &   \underline{p3n}+p4n     &    57.0         &  5.9 $\mu b\ \pm 6.4 \% $    &   4.2 $\mu b$
                 \\
                &   \underline{p5n}+p6n     &    64.0         &  962 $ nb\ \pm 13 \% $       &   1038 $ nb$
                 \\
        & $\alpha$1n+\underline{$\alpha$2n} &    40.0         &  8.8 $\mu b\ \pm 28 \% $     &   1.3 $\mu b$
             \\
        & $\alpha$3n+\underline{$\alpha$4n} &    56.0         &  33. $\mu b\ \pm 17 \% $     &   50.0 $\mu b$
          \\ 
          \\
$ ^{178}Hf$ &     3n   &    39.0         &  1.4 $\mu b\ \pm 13 \% $     &   1.3 $\mu b$
      \\
                &    \underline{4n}+5n      &    47.0         &  3.7 $\mu b\ \pm 6.9 \% $    &   2.96 $\mu b$
                    \\
                &     p2n       &    40.0         &  923 $ nb\ \pm 16 \% $       &   366 $ nb$
                    \\
                &     p3n       &    47.0         &  5.6 $\mu b\ \pm 8.1 \% $    &   3.9 $\mu b$
                    \\
                &   \underline{p4n}+p5n     &    57.0         &  15.7 $\mu b\ \pm 10.3 \% $  &   7.4 $\mu b$
                    \\
        & $\alpha$2n+\underline{$\alpha$3n} &    47.0         &  22.4 $\mu b\ \pm 27 \% $    &   35.1 $\mu b$
            \\
        & \underline{$\alpha$4n}+$\alpha$5n &    56.0         &  61. $\mu b\ \pm 20 \% $     &   70.6 $\mu b$
            \\ 
            \\
$ ^{179}Hf$ &     3n   &    38.0         &  2.1 $\mu b\ \pm 12 \% $     &   1.6 $\mu b$
      \\
                &     4n        &    43.0         &  4.9 $\mu b\ \pm 5.3 \% $    &   11.0 $\mu b$
                \\
                &    \underline{5n}+6n      &    52.0         &  4.6 $\mu b\ \pm 11 \% $     &   4.4 $\mu b$
                \\
                &     p2n       &    41.0         &  352 $ nb\ \pm 31 \% $       &   42 $ nb$
                \\
                &     p3n       &    45.0         &  3.5 $\mu b\ \pm 18 \% $     &   1.1 $\mu b$
                \\
                &     p4n       &    54.0         &  9.2 $\mu b\ \pm 6.1 \% $    &   3.7 $\mu b$
                \\
                &   \underline{p5n}+p6n     &    65.0         &  8.6 $\mu b\ \pm 3.8 \% $    &   4.6 $\mu b$
                \\
        & \underline{$\alpha$3n}+$\alpha$4n &    52.0         &  77.5 $\mu b\ \pm 13 \% $    &   77 $\mu b$
        \\
        & \underline{$\alpha$5n}+$\alpha$6n &    63.0         &  44.3 $\mu b\ \pm 9 \% $     &   93.4 $\mu b$
        \\ 
        \\
$^{180}Hf$ &     3n   &    38.0         &  5.2 $\mu b\ \pm 5.6 \% $    &   0.8 $\mu b$
      \\
                &     4n        &    43.0         &  30.5 $\mu b\ \pm 1.7 \% $   &   17.9 $\mu b$
                \\
                &     5n        &    52.0         &  15.7 $\mu b\ \pm 6 \% $     &   22.5 $\mu b$
                \\
                &    \underline{6n}+7n      &    65.0         &  3.5 $\mu b\ \pm 13 \% $     &   6.1 $\mu b$
                \\
                &     p3n       &    45.0         &  1.9 $\mu b\ \pm 22 \% $     &   2.3 $\mu b$
                \\
                &     p4n       &    54.0         &  15.3 $\mu b\ \pm 6.8 \% $   &   9.6 $\mu b$
                \\
                &     p5n       &    64.0         &  13.1 $\mu b\ \pm 5.7 \% $   &   15.6 $\mu b$
                \\
                & $\alpha 3n$   &    43.0         &  54.8 $\mu b\ \pm 19 \% $    &   16.2 $\mu b$
                \\
        & \underline{$\alpha$4n}+$\alpha$5n &    62.0         &  133 $\mu b\ \pm 8.3 \% $    &   90.2 $\mu b$
            \\ 
            \\
$ ^{181}Ta$ &     3n   &    40.0         &  331 $\mu b\ \pm 25 \% $     &   4.3 $\mu b$
      \\
                &     4n        &    43.0         &  1.1 $\mu b\ \pm 19 \% $     &   42.2 $\mu b$
                \\
                &     5n        &    47.0         &  153 $ nb\ \pm 29 \% $       &   67.4 $\mu b$
                \\
                &     6n        &    61.0         &  71. $ nb\ \pm 43 \% $       &   9.3 $\mu b$
                \\
                &     p3n       &    44.0         &  1.8 $\mu b\ \pm 9.5 \% $    &   0.3 $\mu b$
                \\
                &     p4n       &    54.0         &  1.7 $\mu b\ \pm 5.8 \% $    &   1.0 $\mu b$
                \\
                &   $\alpha$2n  &    36.0         &  1.8 $\mu b\ \pm 32 \% $     &   2.1 $\mu b$
                \\
                &   $\alpha$3n  &    43.0         &  19.6 $\mu b\ \pm 9.8 \% $   &   20.1 $\mu b$
                \\
        & \underline{$\alpha$4n}+$\alpha$5n &    61.0         &  35.5 $\mu b\ \pm 5. \% $    &   43.9 $\mu b$
        \\
        & \underline{$\alpha$6n}+$\alpha$7n &    75.         &  3.1 $\mu b\ \pm 16.2 \% $   &   13.4 $\mu b$
        \\
       \hline \hline
  \end{longtable}

\begin{figure*}
\includegraphics[width=8 cm]{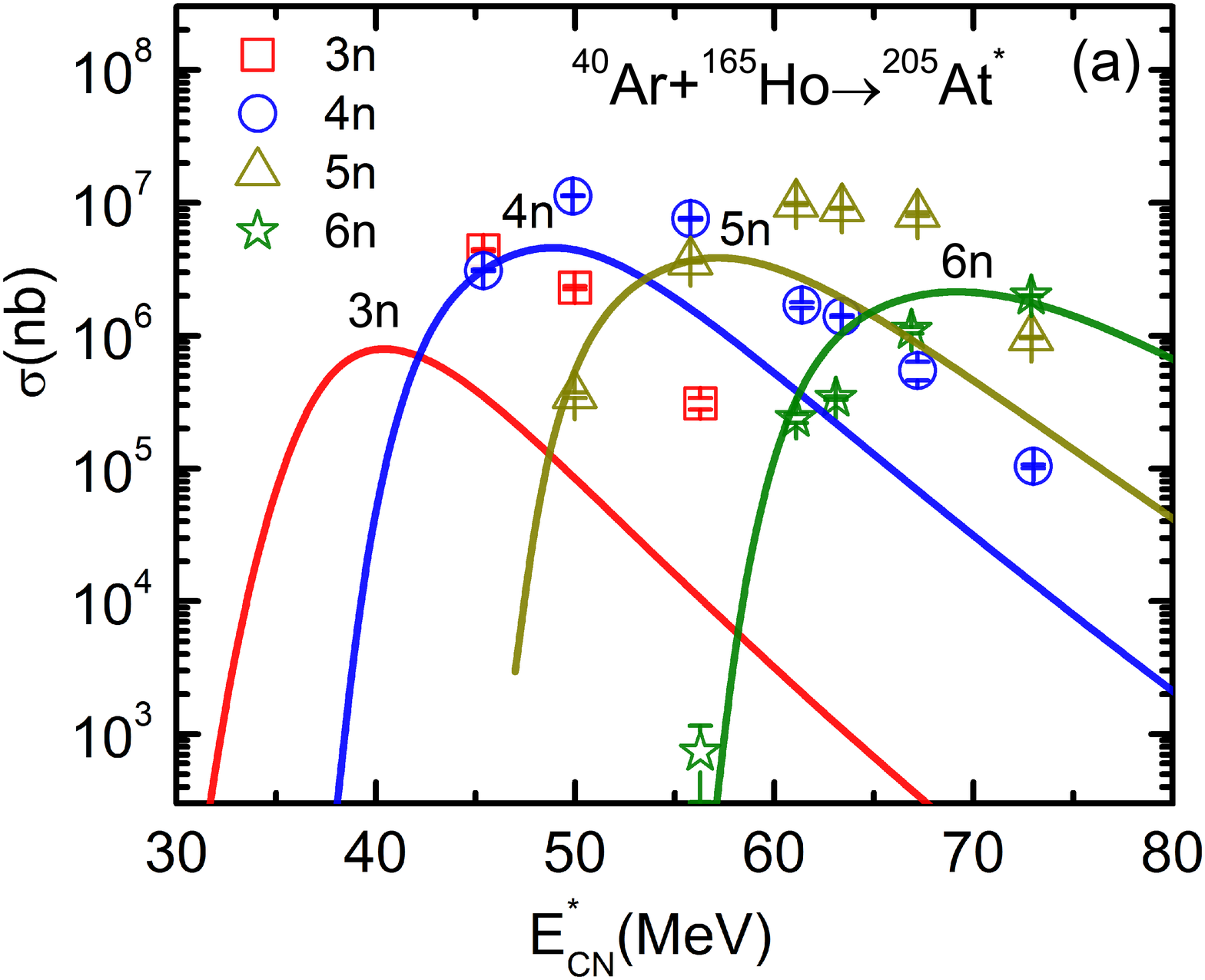}
\includegraphics[width=8 cm]{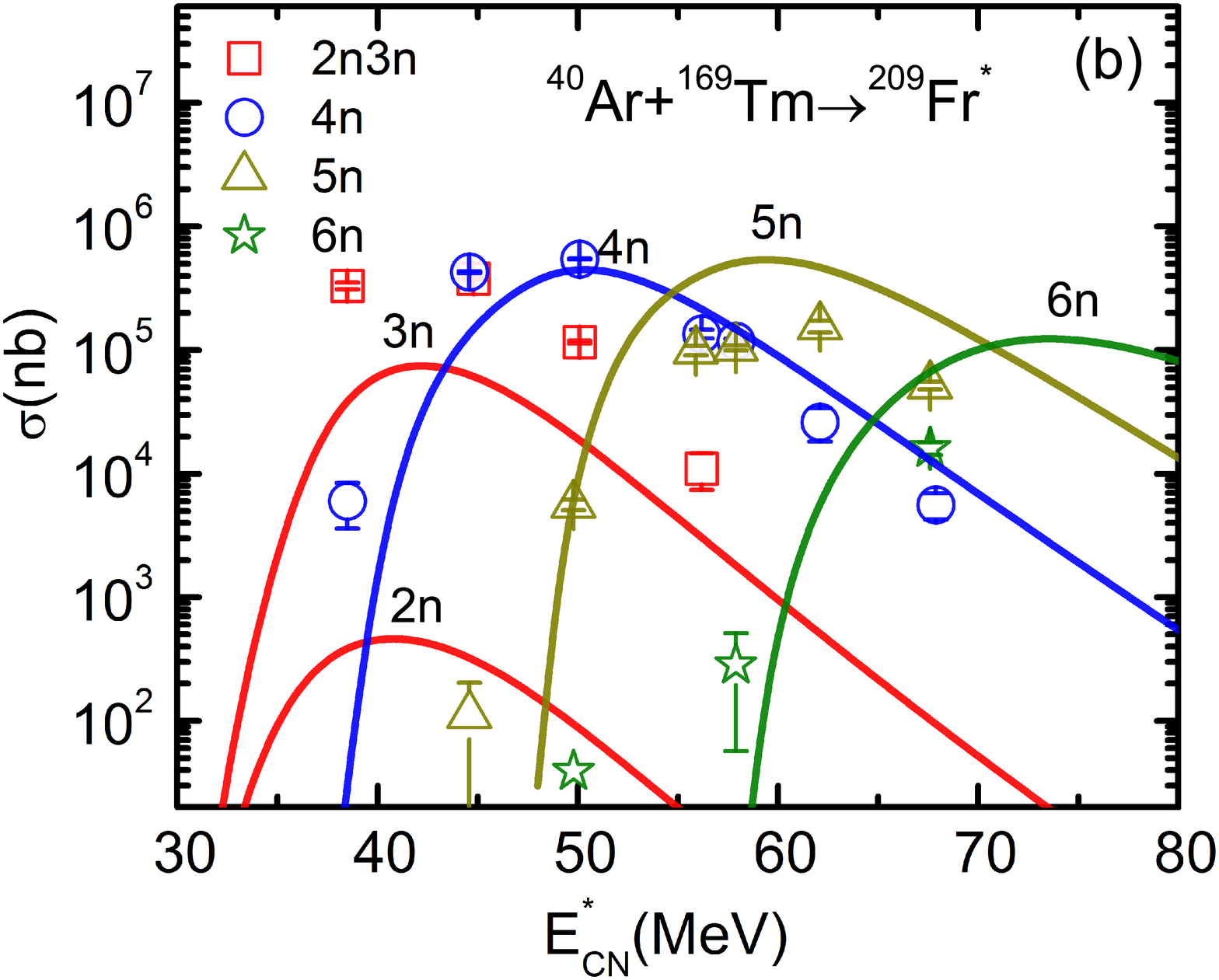}
\caption{\label{fig.1} (Color online) Calculated fusion-evaporation excitation functions and compared with the experimental data in the reactions of $^{40}$Ar + $^{165}$Ho and $^{40}$Ar + $^{169}$Tm \cite{Ve84}. }
\end{figure*}

In order to test the model as discussed in Sec. \Rmnum{2}, the maximal cross sections in the channels stated as shown in Table 1 in the $^{40}$Ar induced fusion-evaporation reactions are calculated and compared with the available data measured at Gesellschaft f\"{u}r Schwerionenforschung at Darmstadt, Germany \cite{Ve84}. The dominant evaporation channels are underlined in the calculation. The charged particle evaporation is comparable to the pure neutron channels for the proton-rich nuclei. The fission barrier is high and even larger than the particle separation energy. It is different to the superheavy region, in which the charged particle evaporation can be neglected in comparison to the neutron emission because of the higher Coulomb barrier \cite{Fe06,Fe09}. Overall, the experimental cross sections can be understood nicely well.

\begin{figure*}
\includegraphics[width=16 cm]{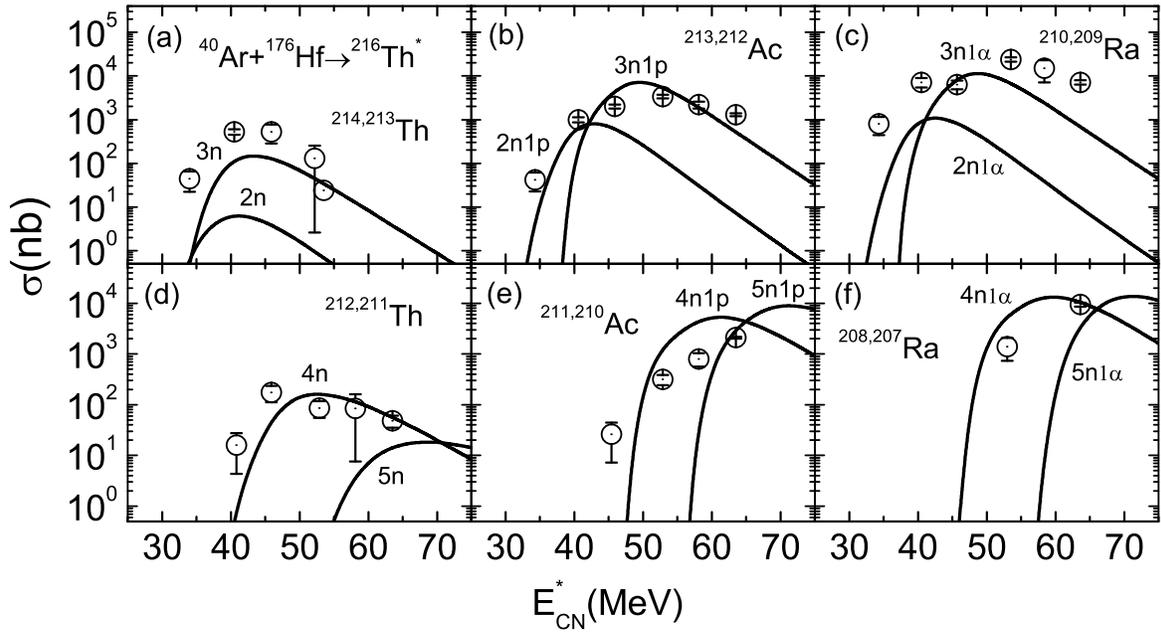}
\caption{\label{fig.2} Similar to in Fig. 1, but for the reaction of $^{40}$Ar + $^{176}$Hf.}
\end{figure*}

\begin{figure*}
\includegraphics[width=16 cm]{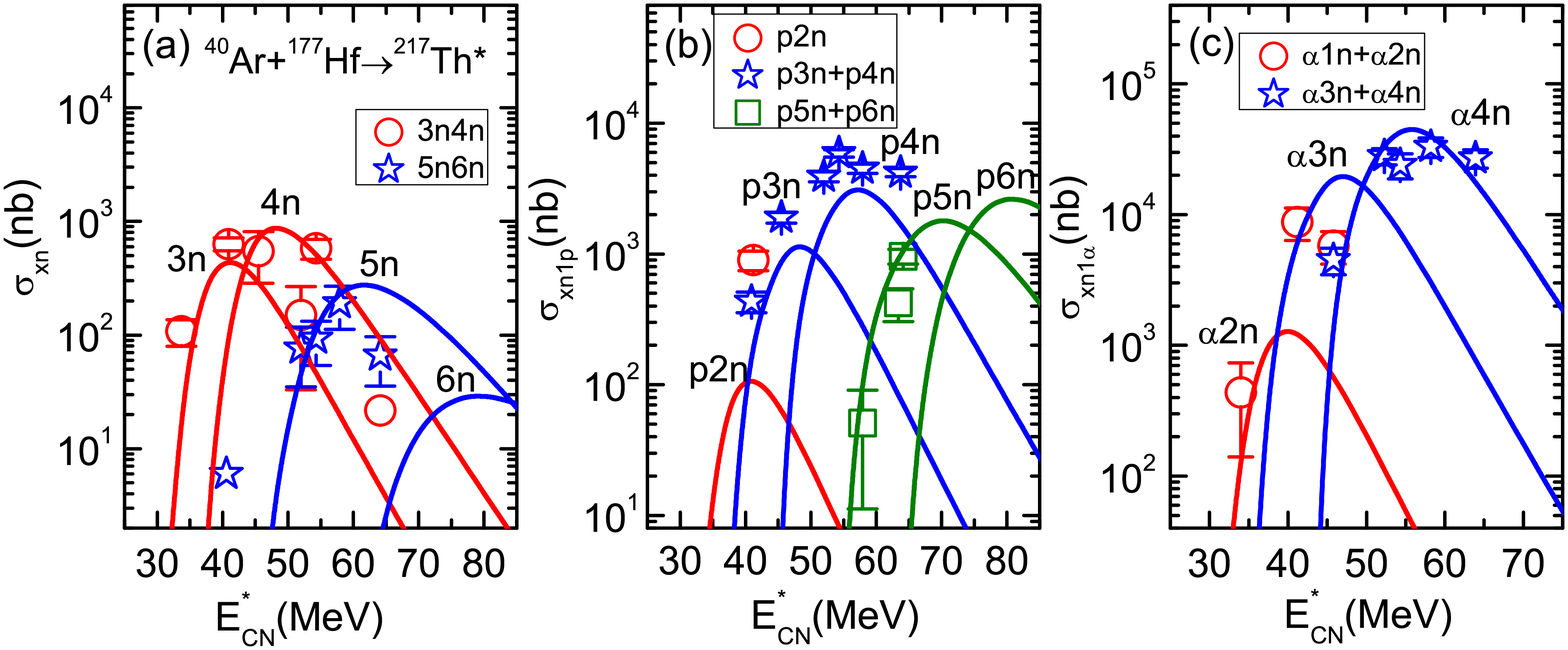}
\includegraphics[width=16 cm]{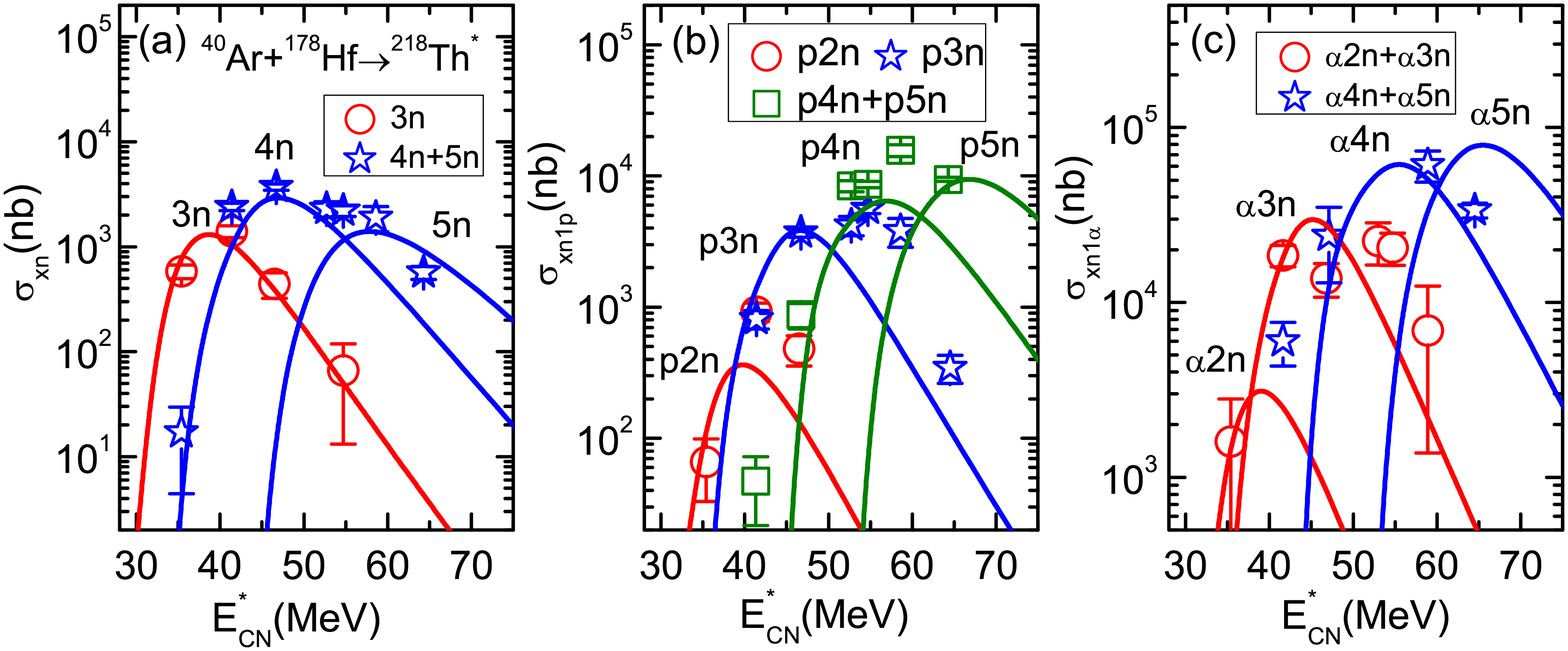}
\caption{\label{fig.3} (Color online) The evaporation residue cross sections with channels of $3-6n$, $1p2-5n$ and $1\alpha2-5n$ in collisions of $^{40}$Ar + $^{177}$Hf (upper panels) and $^{40}$Ar + $^{178}$Hf (lower panels), respectively.}
\end{figure*}

More sophisticated comparison is shown in Fig. 1 for the reactions of $^{40}$Ar + $^{165}$Ho (left panel) and $^{40}$Ar + $^{169}$Tm (right panel). The evaporation channels are labeled with the $2-6n$ de-excitation from the compound nucleus. The mass table and shell correction energy from Ref. \cite{Mo95} are taken in the calculation. Different to the superheavy nucleus formation, the maximal cross sections of the $4-6n$ channels are similar because the fission barrier is not reduced with increasing the excitation energy. The neutron numbers (120 and 122) of the excited compound nuclei are below the shell closure N=126, which lead to the neutron evaporation dominating the decay process. However, the neutron evaporation is strongly suppressed for the neutron number around the shell closure as shown in Fig. 2 and Fig. 3 in collisions of $^{40}$Ar on $^{176}$Hf and $^{177,178}$Hf, respectively. The combined channels with proton or $\alpha$ contribute the cooling process. The channels of $1p4n$, $1p5n$, $1\alpha4n$ and $1\alpha5n$ have larger cross sections than the $4-6n$ evaporation. The cross sections of pure neutron evaporation increase rapidly with the neutron numbers of compound nuclei as shown in Fig. 4. However, the channels with charged particles weakly depend on the isotopic targets. The maximal cross sections of charged particle evaporation are similar because the fission barrier does not decrease after emitting a neutron, e.g., $1p4-6n$, $1\alpha4-6n$. The available data are nicely reproduced with the DNS model and each evaporation channels can be clearly explained.

\begin{figure*}
\includegraphics[width=16 cm]{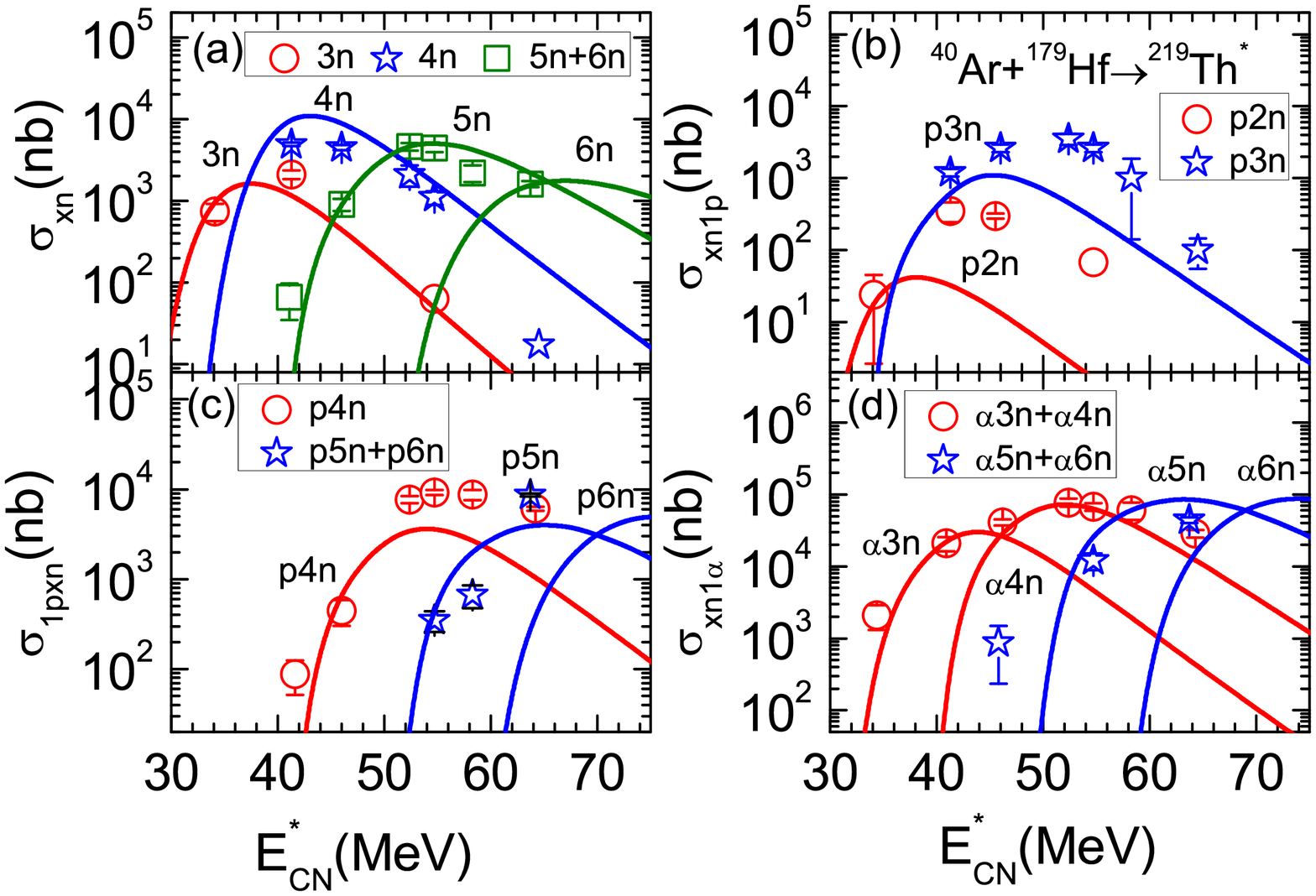}
\includegraphics[width=16 cm]{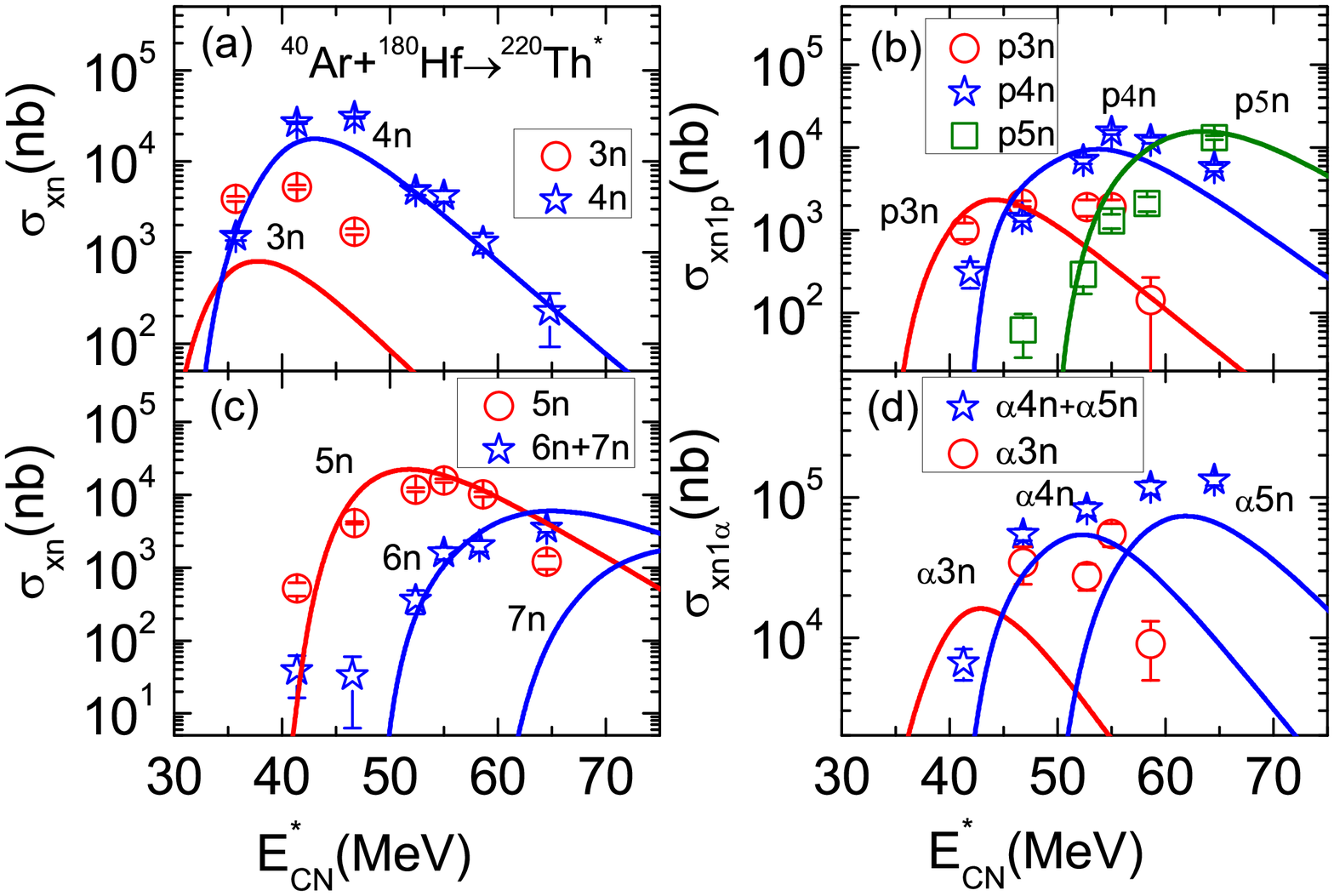}
\caption{\label{fig.4} (Color online) The same as Fig. 3, but for on the isotopic targets of $^{179}$Hf and $^{180}$Hf, respectively.}
\end{figure*}

\subsection{Projectile dependence in the production of proton rich nuclei around Z=84-90}

\begin{figure*}
\includegraphics[width=15 cm]{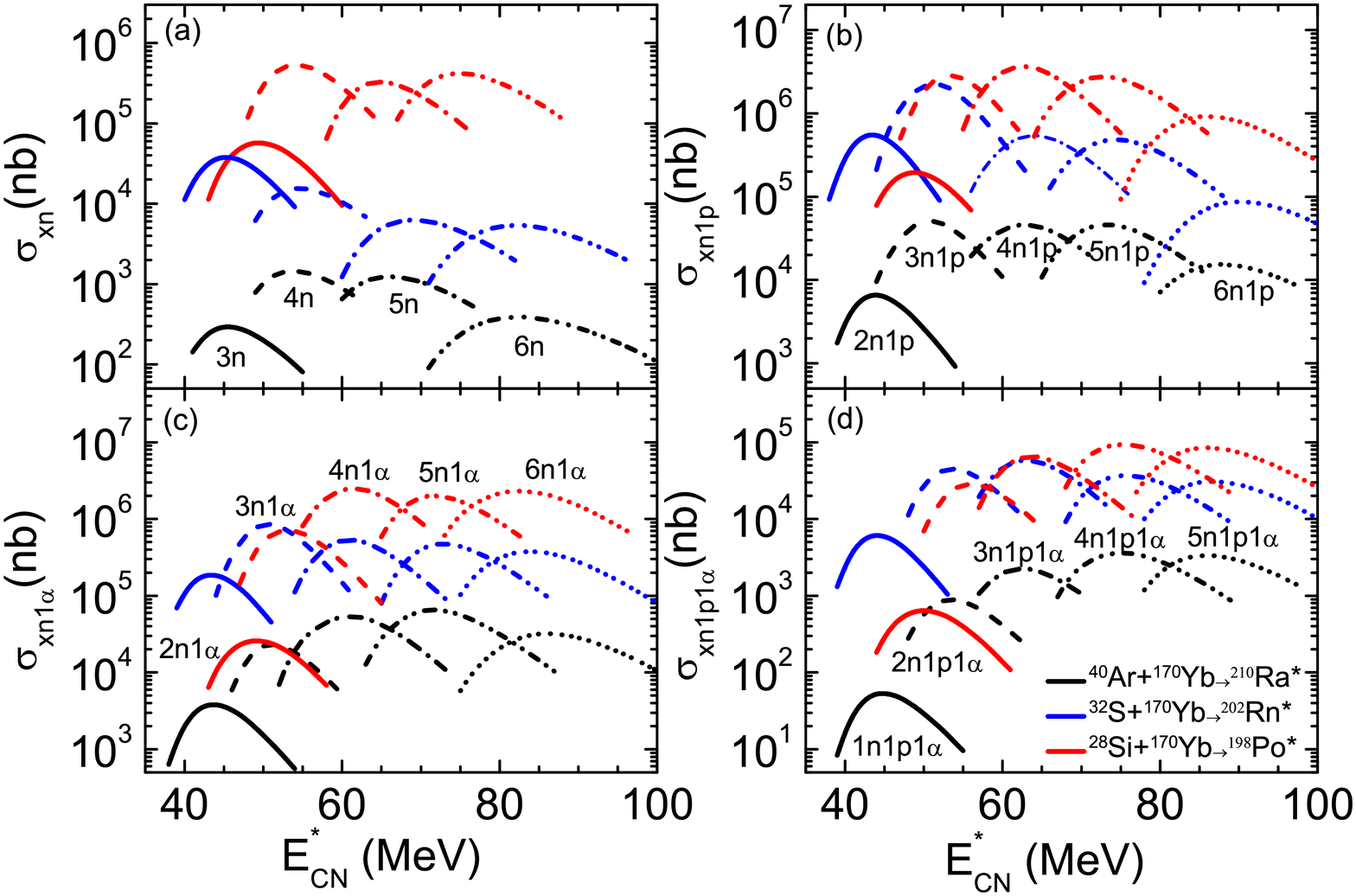}
\caption{\label{fig.5} (Color online) The fusion-evaporation excitation functions in collisions of $^{28}$Si, $^{32}$S and
$^{40}$Ar on $^{170}$Yb, respectively.}
\end{figure*}

\begin{figure*}
\includegraphics[width=15 cm]{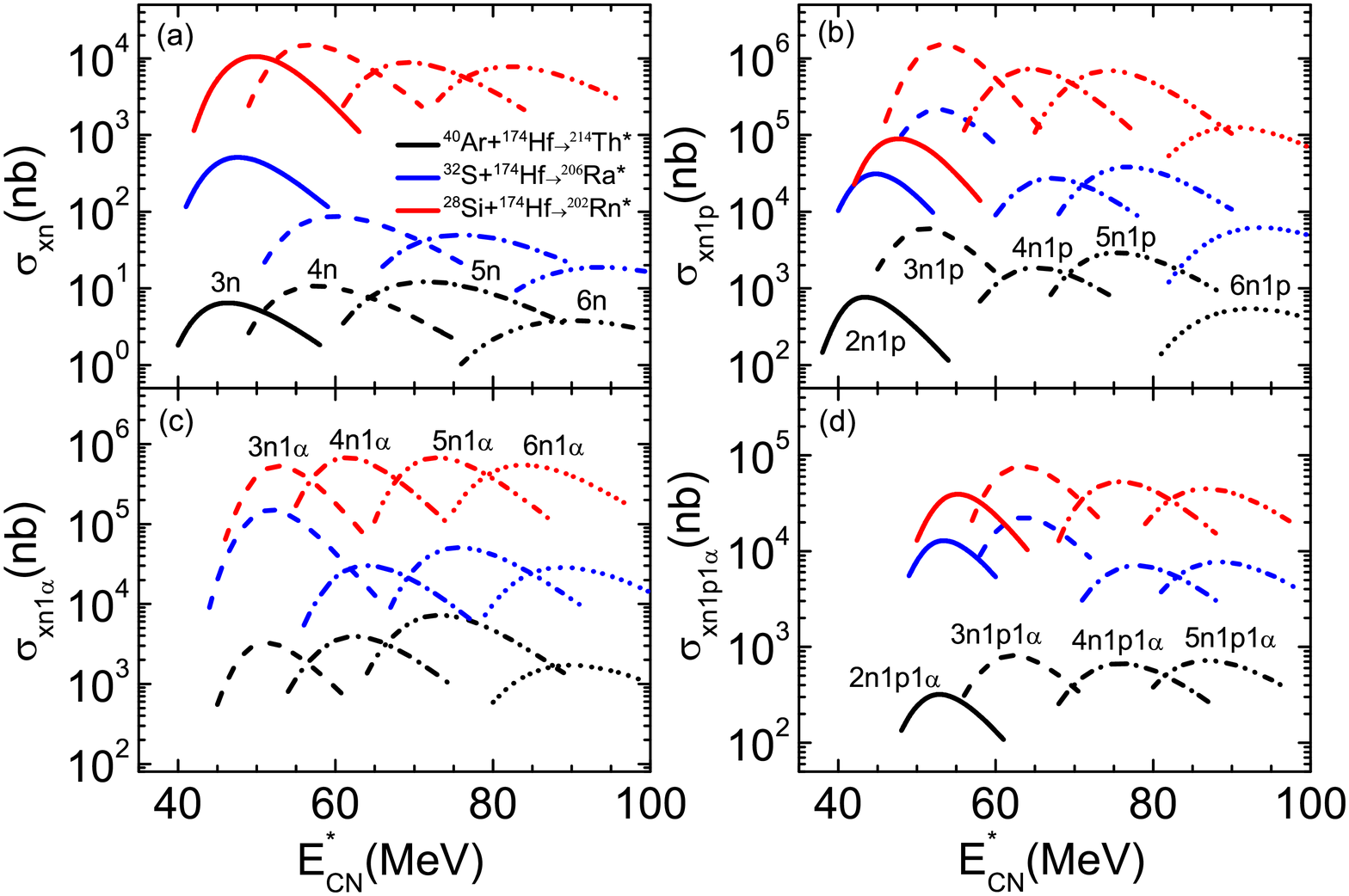}
\caption{\label{fig.6} (Color online) The same as Fig. 5, but on the target of $^{174}$Hf.}
\end{figure*}

The influence of entrance systems on the formation of proton-rich nuclei is analyzed thoroughly within the DNS model. Shown in Fig. 5 is the fusion-evaporation cross sections with $^{28}$Si, $^{32}$S and $^{40}$Ar bombarding $^{170}$Yb for producing the neutron-rich around Z=83-88 from the compound nuclei $^{198}$Po$^*$, $^{202}$Rn$^*$ and $^{210}$Ra$^*$. The red, blue and black lines correspond to the projectiles of $^{28}$Si, $^{32}$S and $^{40}$Ar, respectively. The different symbols from left to right side in each panel represent the evaporation channels, e.g., $3-6n$, $2-6n1p$, $2-6n1\alpha$ and $1-5n1p1\alpha$ respective to the panels of (a), (b), (c) and (d) respectively. It should be noticed that the residue cross sections increase with the mass asymmetry because of the reduction of the inner fusion barrier except for the left channels, i.e., $3n$, $2n1p$, $2n1\alpha$ and $1n1p1\alpha$, which are attributed from the competition of fusion and survival probabilities. The even-odd effect is pronounced in the pure neutron evaporation. However, the maximal cross sections are similar in the charged particle channels, e.g., $3-5n1p$, $4-6n1\alpha$ and $3-5n1p1\alpha$. The fission barriers are larger than the particle separation energy in the region of Z=82-90 \cite{Mo09}. The fusion probability reaches a constant value at the excitation energies above 40 MeV. Therefore, the maximal cross sections for the charged channels are not reduced with increasing the excitation energy. The fusion-evaporation excitation functions are further analyzed in Fig. 6, Fig. 7 and Fig. 8 for the targets of $^{174}$Hf, $^{175}$Lu and $^{181}$Ta, respectively. Besides the inner fusion barrier of DNS, the particle separation energy is also important on the residue cross section, in particular for the neutron evaporation around the shell closure N=126. It is obvious that the neutron channels for the targets of $^{170}$Yb and $^{174}$Hf have smaller cross sections in comparison to $^{175}$Lu and $^{181}$Ta.

\begin{figure*}
\includegraphics[width=15 cm]{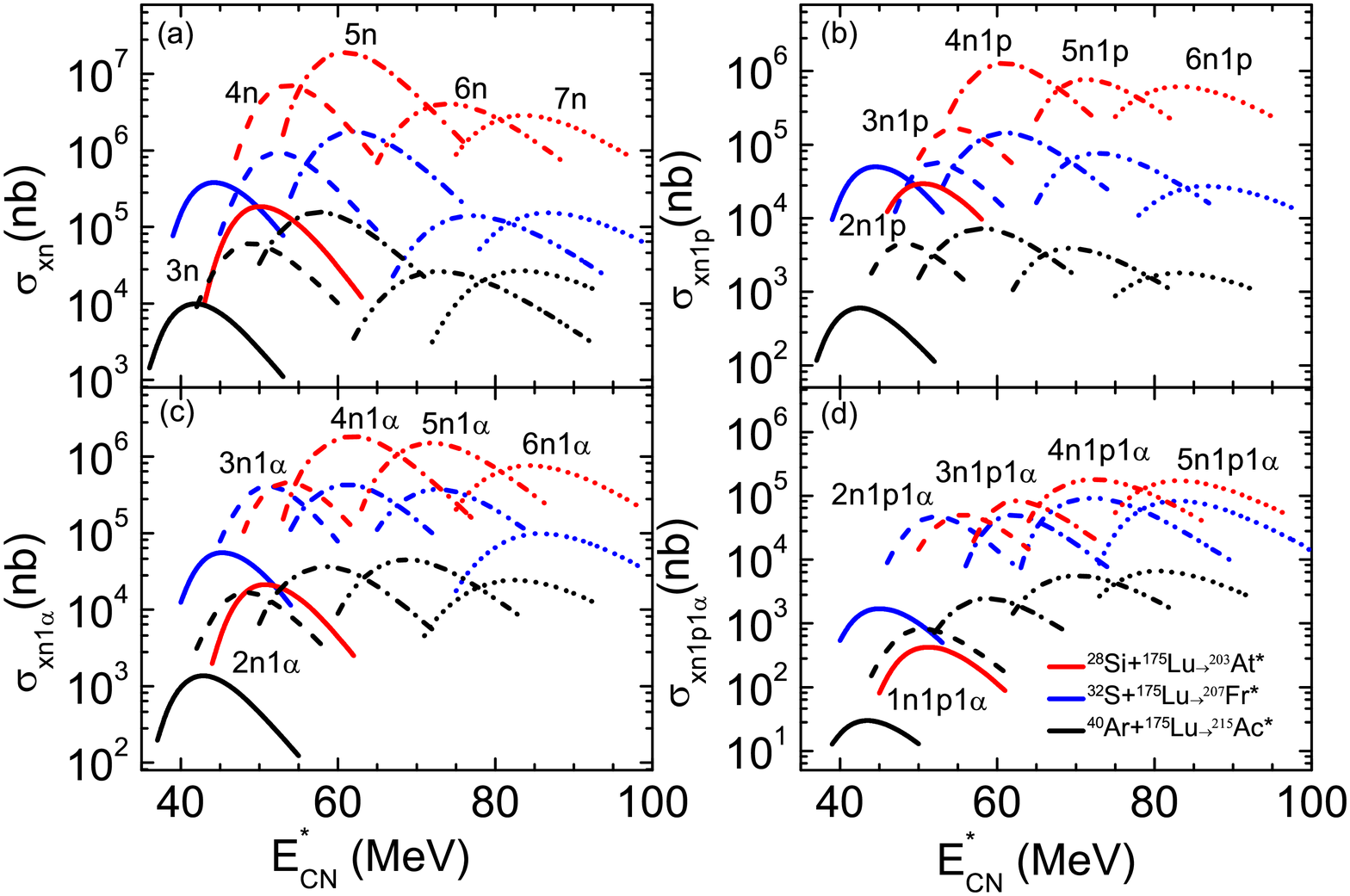}
\caption{\label{fig.7} (Color online) The evaporation residue cross sections in the channels of $3-7n$, $1p2-6n$, $1\alpha2-6n$ and $1p1\alpha2-5n$ in the reactions of $^{28}$Si, $^{32}$S, $^{40}$Ar + $^{175}$Lu, respectively.}
\end{figure*}

\begin{figure*}
\includegraphics[width=15 cm]{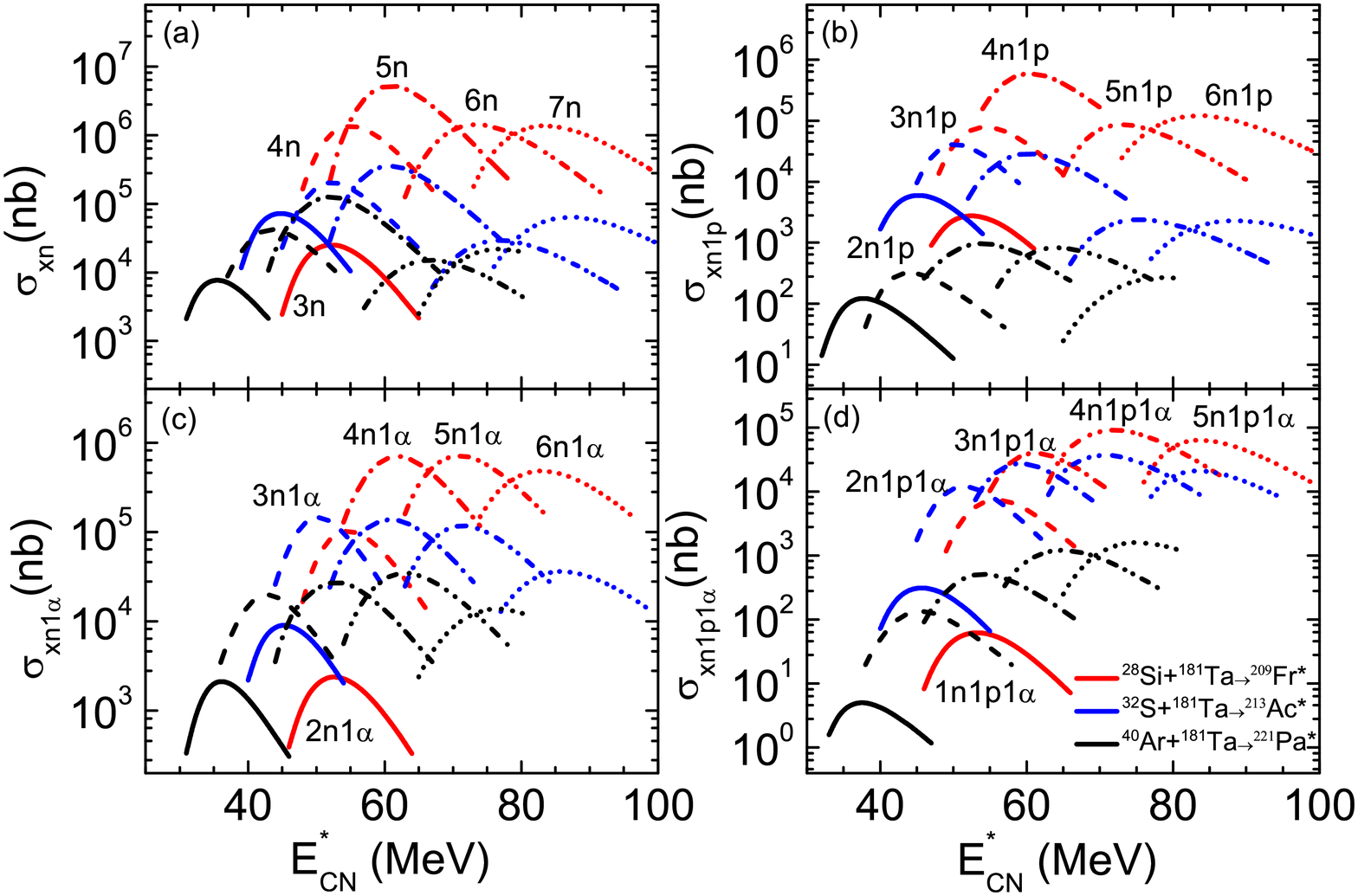}
\caption{\label{fig.8} (Color online) The same as Fig. 7, but for the target of $^{181}$Ta.}
\end{figure*}

The reactions of $^{28}$Si + $^{174}$Hf and $^{32}$S + $^{170}$Yb lead to the same compound nucleus $^{202}$Rn formation. The both systems have the similar capture and fusion probabilities at the same excitation energy, which result in the same structure of the evaporation residue excitations in the channels of $xn$, $xn1p$, $xn1\alpha$ and $xn1p1\alpha$. For the production of Fr (Z=87) isotopes, the neutron evaporation channels $xn$ of reactions $^{28}$Si + $^{181}$Ta and $^{32}$S + $^{175}$Lu, in $xn1p$ channels of reactions $^{32}$S + $^{174}$Hf and $^{40}$Ar + $^{170}$Yb, the $xn1\alpha$ channels of reactions $^{32}$S + $^{181}$Ta and $^{40}$Ar + $^{175}$Lu and $xn1p1\alpha$ channels of reactions $^{40}$Ar + $^{174}$Hf are possible with the maximal cross sections of $5\times10^6$ nb, $4\times10^4$ nb, $4\times10^4$ nb, $2\times10^5$, $4\times10^5$ and $10^3$ nb, respectively. Therefore, the reaction of $^{28}$Si + $^{181}$Ta in the neutron evaporation channels is proposed for the proton-rich Fr isotopes, in particular for $^{204}$Fr. For the production of Ra (Z=88) proton-rich isotopes, the neutron channels of $^{32}$S + $^{174}$Hf and $^{40}$Ar + $^{170}$Yb, the charged particle channels $xn1\alpha$ of $^{40}$Ar + $^{174}$Hf and $xn1p1\alpha$ channels of $^{40}$Ar + $^{181}$Ta are feasible with the maximal cross sections of $10^2$ nb, $10^3$ nb, $10^4$ nb and $10^5$ nb, respectively. So the best combination for producing Ra isotopes is the $^{40}$Ar + $^{181}$Ta via $xn1p1\alpha$ evaporation channels, in which the residue nuclei are created around the neutron shell closure of N=126. The neutron channels of $^{32}$S + $^{181}$Ta and $^{40}$Ar + $^{175}$Lu, the $xn1p$ channels of $^{40}$Ar + $^{174}$Hf and the $xn1\alpha$ channels of $^{40}$Ar + $^{181}$Ta are available for the production of Ac (Z=89) isotopes with the maximal cross sections of $3\times10^5$ nb, $10^5$ nb, $6\times10^3$ and $4\times10^4$ nb, respectively. It is obvious that the systems of $^{32}$S + $^{181}$Ta and $^{40}$Ar + $^{175}$Lu in the neutron evaporation are favorable for producing the proton-rich Ac production. The optimal projectile-target combinations in producing the proton-rich nuclei around Z=84-90 are the competition of the pure neutron evaporation and charged particle emissions.

\subsection{Isotopic dependence on production cross section of proton-rich nuclei}

The production of PRN depends on the isotopic combination of the target and projectile in the fusion-evaporation reactions. For example, the maximal cross section is $174\pm62.64$ nb for the reaction of $^{40}$Ar + $^{176}$Hf $\rightarrow$ $^{212}$Th + $4n$. However, it becomes the value of $30.5\pm0.52$ $\mu$b for the reaction $^{40}$Ar + $^{180}$Hf $\rightarrow$ $^{216}$Th + $4n$ \cite{Ve84}. Studies of isotopic trends in the production PRN would be helpful for predicting the optimal combinations and the optimal excitation energies (incident energy) in experiments. Shown in Fig. 9 and in Fig. 10 is the isotopic distributions with the targets of Yb and Hf, respectively. The channels of $4-6n$, $3n1\alpha$, $4n1\alpha$ and $4n1p$ are analyzed thoroughly with the projectiles of $^{28}$Si, $^{32}$S and $^{40}$Ar, respectively. The available experimental data (circles with error bars) in the reactions of $^{40}$Ar + $^{171,174}$Yb and $^{40}$Ar + $^{176-180}$Hf are shown for comparison. It is obvious
that the neutron evaporation cross section increases with the mass of isotopic target because of the reduction of separation energy. However, the channels with the charged particles are independent on the mass number of targets, but related to the projectile-target asymmetry.

The isotopic trends are mainly caused from the inner fusion barrier of DNS, particle separation energy and fission barrier. When the target neutrons increase, the DNS gets more asymmetry and the fusion probability is enhanced owing to the reduction of inner fusion barrier. A smaller neutron separation energy and a higher fission barrier lead to a larger survival probability. The nucleus around shell closure has a larger shell correction energy and neutron separation energy. For the PRN around Z=84-90, the fission barrier is mainly contributed from the macroscopic energy (The first term in Eq. (33)) and even larger than the particle separation energy. The charged particle escaped from the mother nucleus is comparable to neutron evaporation for the proton-rich nuclei. The residue nucleus becomes more neutron-rich after emitting a charged particle. The experimental data in the $^{40}$Ar + $^{171,174}$Yb reactions are underestimated with the DNS model. But the isotopic structure is consistent with the calculations. All channels of the $^{40}$Ar + $^{176-180}$Hf reactions are nicely reproduced. The systems could be easily constructed in experiments. The attempts to explore the structure information and decay properties of the proton-rich nuclei are proposed, in particular at the Heavy-Ion Accelerator Facility in Lanzhou (HIRFL).

\begin{figure*}
\includegraphics[width=18 cm]{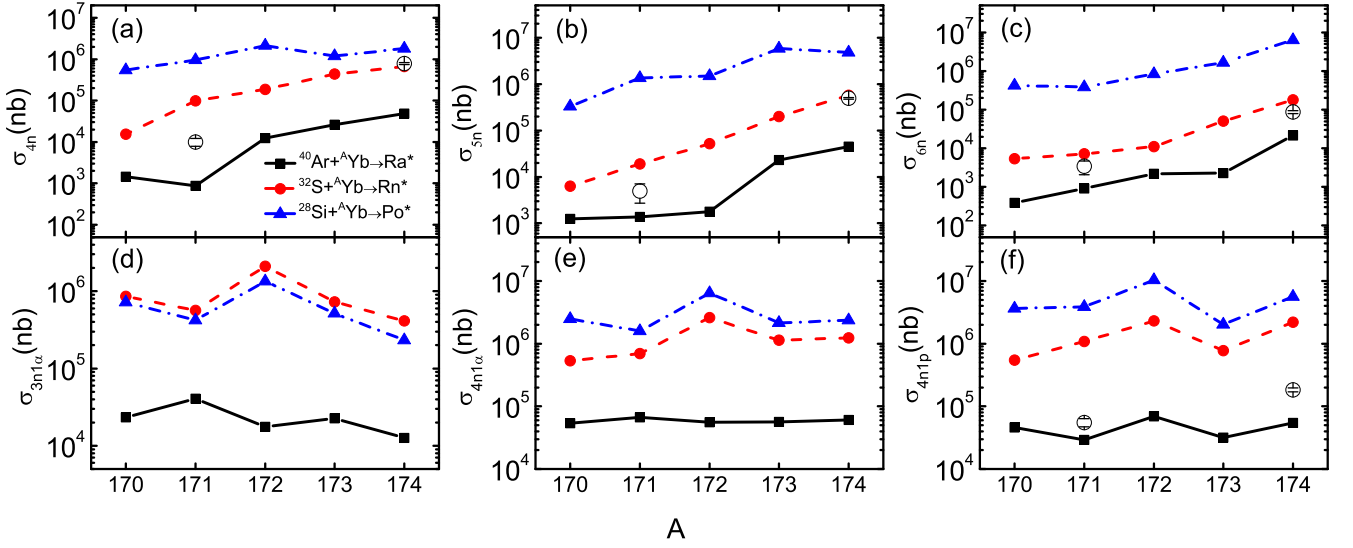}
\caption{\label{fig.9} (Color online) Production cross sections in the evaporation channels of $4-6n$, $1\alpha3n$, $1\alpha4n$ and $1p4n$ as a function of mass number of Yb (Z=70) isotopes in the $^{28}$Si, $^{32}$S and $^{40}$Ar induced reactions.}
\end{figure*}

\begin{figure*}
\includegraphics[width=18 cm]{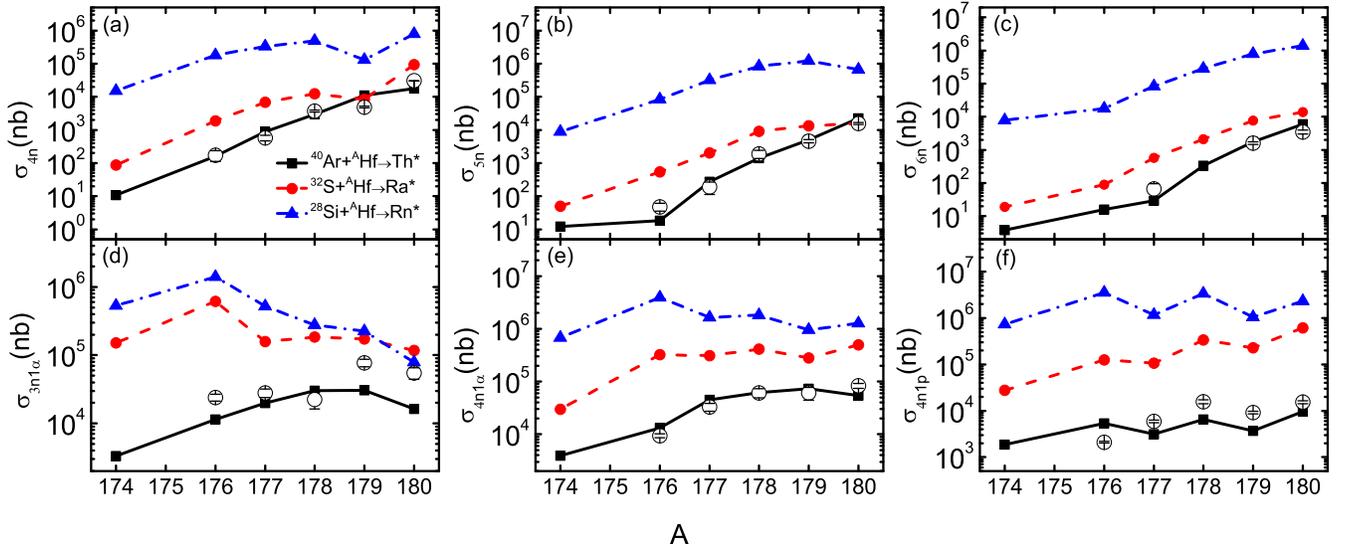}
\caption{\label{fig.10} (Color online) Isotopic dependence in the production of proton-rich nuclei in the domain of Z=84-90.}
\end{figure*}

\section{Conclusions}

Within the DNS concept, a dynamical model is used for describing the production of proton-rich nuclei in fusion-evaporation reactions. The fusion dynamics in the hot fusion reactions for producing the proton-rich nuclei around Z=84-90 are investigated systematically. The calculated results are in good agreement with available experimental data within error bars. The neutron shell closure of N=126 continues to be of importance in the proton-rich domain. The odd-even effect appears in the neutron evaporation. The residue cross sections in the neutron channels increases with the mass of isotopic target in the $^{28}$Si, $^{32}$S and $^{40}$Ar induced reactions. However, the channels with the charged particles are independent on the mass number of targets, but related to the projectile-target asymmetry. The maximal cross sections in the channels of $4-7n$, $1\alpha3-6n$, $1p3-6n$ and $1\alpha1p3-5n$ for the PRN production weakly depend on the excitation energy.

The physical nature of the synthesis of heavy fissile nuclei in fusion-evaporation reactions is very complicated, involving not only certain quantities that crucially influence the whole process but also the dynamics of the process. The coupling of the dynamic deformation and the nucleon transfer in the course of overcoming the multidimensional PES has to be considered in the DNS model. The height of the fission barrier for heavy nuclei is mainly determined by the shell correction energies at the ground state and at the saddle point. Further work is in progress.

\section{Acknowledgements}

We would like to thank Tian-Heng Huang, Zai-Guo Gan, Zhong Liu and Xiao-Hong Zhou for fruitful discussions and experimental possibilities at HIRFL. This work was supported by the Major State Basic Research Development Program in China (Grant Nos 2014CB845405 and 2015CB856903), the National Natural Science Foundation of China (Grant Nos 11675226, 11175218, 11675066 and U1332207), and the Youth Innovation Promotion Association of Chinese Academy of Sciences.

\end{document}